%% file: inequality3.tex
\documentclass[10pt]{article}

\usepackage[letterpaper, margin=2.5cm]{geometry}
\usepackage{amsmath, amssymb, amsthm}
\newtheorem{proposition}{Proposition}
\newtheorem{remark}{Remark}
\usepackage{setspace}
\usepackage{graphicx}
\usepackage{subcaption}
\usepackage{booktabs}
\usepackage{cpblTables}
\graphicspath{{./}}
\renewcommand\ctDraftComment[1]{} 
\usepackage[backend=biber, natbib=true, style=authoryear,
            maxcitenames=2, maxbibnames=99,
            uniquelist=false, uniquename=false]{biblatex}
\usepackage{microtype}
\usepackage{xspace}

\usepackage{refstyle}
\newref{tab}{name={},Name={},
  refcmd={\hyperref[#1]{\ifRScapname Table\else table\fi~\ref{#1}}}}
\newref{fig}{name={},Name={},
  refcmd={\hyperref[#1]{\ifRScapname Figure\else figure\fi~\ref{#1}}}}
\newref{sec}{name={},Name={},
  refcmd={\hyperref[#1]{\ifRScapname Section\else section\fi~\ref{#1}}}}
\newref{app}{name={},Name={},
  refcmd={\hyperref[#1]{\ifRScapname Appendix\else appendix\fi~\ref{#1}}}}
\newref{prop}{name={},Name={},
  refcmd={\hyperref[#1]{\ifRScapname Proposition\else proposition\fi~\ref{#1}}}}
\newref{eq}{name={},Name={},
  refcmd={\hyperref[#1]{Equation~\ref{#1}}}}

\newcommand{\fvset}{\{0,5,10\}}
\DeclareRobustCommand{\FVR}{{\fontseries{m}\scshape fvr}} 
\newcommand{\FVRI}{\textsc{fvri}}

\newcommand{\SWL}{life satisfaction}
\newcommand{\Swl}{Life satisfaction}
\newcommand{\GHM}{GHM\xspace}%
\newcommand{\CF}{Cowell--Flachaire}
\newcommand{\sd}{\sigma}
\newcommand{\sdobs}{\sigma^{\mathrm{obs}}}
\newcommand{\sdtrue}{\sigma^{*}}
\newcommand{\sdpred}{\hat\sigma^{\mathrm{pred}}}

\newcommand{\latestversionline}{Please use
  \href{https://alum.mit.edu/www/cpbl/publications/Barrington-Leigh-JWELL2026-SWBinequality.pdf}{the
    latest version of this paper online}.}

\ifdefined\blindreview\fi

\newcommand{\specB}{(B)}%

\usepackage[colorlinks=true, linkcolor=blue, citecolor=blue, urlcolor=blue]{hyperref}

\usepackage[user, abspage]{zref}

\title{Does life-satisfaction inequality measure societal inequality?\\
  { A focal-value-rounding critique}}

  \author{C.P.~Barrington-Leigh\thanks{
      Author contact at \url{https://alum.mit.edu/www/cpbl/contact}.
      Comments welcome.\\
  \href{https://alum.mit.edu/www/cpbl/publications/SWBinequality/}{
    Interactive demonstrations available online}.
  AI was used to support this research.
  This work was supported by SSHRC grant 435-2024-1219. I am grateful to
  Gaston Yalonetzky, audiences in Vancouver and Luxembourg, and two referees
  for helpful comments.
  \protect \\
  }}
  \date{{\fontsize{2}{3}\selectfont v2.0 : compiled 4 August 2026.   First draft: Solstice 2026} \\
  \latestversionline
  }

\begin{document}
\maketitle

\begin{abstract}
The dispersion of self-reported \SWL{} has been proposed
and used as a comprehensive measure of societal inequality. A negative
cross-country association between mean \SWL{} and its standard deviation
has been read as evidence that this inequality is itself welfare-relevant, but critics have pointed to the nonlinearity and boundedness of response
scales.
I describe a further problem: a substantial and
predictable share of respondents simplify the 0--10 scale to the 
subset $\fvset$ --- ``focal-value rounding'' (\FVR) --- a behaviour that
breaks not only linearity of the scale but even its order,
in the sense that no stretching or relabelling of response categories can undo \FVR{}.
I derive a sharp bound
on the bias 
that \FVR{}
can induce in the
standard-deviation;
at empirically-typical \FVR{} fractions 
the possible bias is roughly half the
cross-country range of  dispersion.
Turning to empirical data, I estimate and correct for the 
bias by fitting a model of \FVR{} behavior
to the Gallup World
Poll Cantril ladder (135 countries;
$N=300{,}137$) and three other multi-country life evaluations.
The 
corrections are material: \FVR{} inflates measured standard deviation
in 133 of 135
countries, by a median of 0.090 points. However, because the corrections
are nearly uniform in sign across countries,
rankings of the standard deviation survive essentially intact.
The ordinal inequality indices which \citet{CowellFlachaire2017}
advocate as the theoretically appropriate alternative
suffer even worse from \FVR{}.
The correlation between mean and standard deviation \citep{Goff-Helliwell-Mayraz-EI2018-SWB-inequality}
survives \FVR{} correction with modest
attenuation and some caveats. Thus \FVR{} does not, by itself, overturn previously reported
correlations;
it does, however, add
a further reason for caution in interpreting scalar measures of subjective wellbeing inequality: candidate statistics are differently
contaminated, the ordinal repairs no less than the cardinal originals, and they
disagree with one another.
Lastly, I show that \FVR{}-correction can improve the pairwise dominance-comparability of full distributions.

\end{abstract}


\newpage
\tableofcontents
\newpage
\listoffigures
\listoftables
\newpage

\section{Introduction}\label{sec:intro}

If self-reported \SWL{} is accepted as a comprehensive welfare measure ---
one that captures health, relationships, security, and purpose, and other human experience in appropriate measure along with
income --- then its distribution within a population is of as much
interest as its mean.
Accordingly, researchers have computed
inequality statistics directly from the distribution of \SWL{} responses. Cross-national comparison of happiness
dispersion dates at least to \citet{Veenhoven-1990-inequality-in-happiness},
and a methodological literature has debated which statistic is proper to the
task at least since \citet{springerlink:10.1007/s10902-005-8855-7}, with
\citet{DelheyKohler2011} proposing a modified  standard deviation,
\citet{Hasegawa-Ueda-JSE2011-SWB-inequality} proposing an inequality measure they call ``regret,''
and
\citet{Berenger-Silber-JOHS2022-measurement-happiness-inequality} reviewing
the now-substantial menu of candidate measures. Within this lineage,
\citet[][hereafter, \GHM{}]{Goff-Helliwell-Mayraz-EI2018-SWB-inequality} use the within-country
standard deviation; \citet{Stevenson-Wolfers-NBER2008-happiness-inequality-United-States}
use the Gini coefficient; and practitioner guidance recommends a portfolio
including the standard deviation, threshold shares, the Gini, and the
Shannon index \citep{WWCW2018}.
The choice among
distribution-sensitive aggregates inevitably encodes an ethical stance, and
\citet{PanasiukMcCannyCheung2025} show that \SWL{} responses depart from
normality in every one of 148 countries --- so that no mean-plus-dispersion
summary captures the distribution.

The stakes have been raised by a substantive empirical claim:
\citet{Goff-Helliwell-Mayraz-EI2018-SWB-inequality} report that
country-mean \SWL{} is lower where the standard deviation of \SWL{} is
higher, conditional on income, and read this as evidence that wellbeing
inequality is itself welfare-reducing, and more so than income
inequality. Two claims are therefore under test when a dispersion
statistic of \SWL{} is reported: that cross-country variation in the
statistic meaningfully reflects variation in some underlying wellbeing inequality; and
that this inequality is welfare-relevant in the way the mean--dispersion
association suggests.

Doubt about the first claim is natural given the analogous concern about  cardinality of the mean, though the latter issue turns out to have limited empirical impact
 \citep{Ferrer-i-Carbonell-Frijters-EJ2004,Kaiser-Vendrik-DRAFT2020-debunking-Bond-Lang}.
An ordinal-data literature has produced, most recently, the status-based
index family of \citet{CowellFlachaire2017} and its associated dominance
theory \citep{Jenkins2020-RIW-ordinal, GravelMagdalouMoyes2020}.
\citet{Grimes-Jenkins-Tranquilli-JoHS2023-SWB-inequality} bring this
machinery to the \GHM{} question and find that the negative
mean--dispersion association is not robust: it vanishes when country and
wave fixed effects are included, and the theoretically appropriate ordinal
indices yield relationships of opposite signs depending on whether status
is computed looking upward or downward.

This paper describes a further measurement problem, distinct from
ordinality, and quantifies its consequences for both claims.
\citet{Barrington-Leigh-JPubEcon2024-focal-values} documents that a
substantial share of respondents simplify a 0--10 response scale to the
focal subset $\fvset$, and that the propensity to do so is predictable ---
most strongly by education. This ``focal-value rounding'' (\FVR, a form of response heaping in subjective numeric data) breaks not just
the cardinality of the reported scale but its ordinality: a respondent who
maps an interior assessment to 5 produces a report that no monotone
transformation of the labels can undo.  \FVR{} of this kind is plainly visible in published
cross-country response histograms \citep[e.g., Figure 4 of][]{Grimes-Jenkins-Tranquilli-JoHS2023-SWB-inequality}, though never modeled in the
literature on \SWL{} inequality.

This paper makes four contributions. First, theory: I derive a bound on the
bias that \FVR{} can induce in the standard deviation of reported \SWL{}
--- the dispersion analogue of the mean-bias bound in
\citet[Proposition~1]{Barrington-Leigh-JPubEcon2024-focal-values} --- and
show in simulation that the bias can take either sign, that it varies with
the latent mean, and that it propagates undiminished into the \CF{}
ordinal indices. 
The reason is that those indices are functionals of the response CDF, which is necessarily changed by \FVR{};
an index immune to scale-stretching is not \FVR{}-proof.
In a synthetic cross-section of countries with no genuine wellbeing-inequality
channel (latent dispersion identical across countries by construction),
heterogeneous \FVR{} rates
nonetheless contaminate the cross-country mean--dispersion comparison,
spuriously reproduce \GHM's inequality-aversion interaction for the upper range of the \SWL{} scale, and contaminate
the ordinal variation-ratio measure.

Second, measurement: I adapt the model of
\citet{Barrington-Leigh-JPubEcon2024-focal-values} to allow for separate, predictable
rounding propensities toward each focal value, layered on an ordered
probit with a hierarchical per-country latent scale. I fit this to four
large multi-country datasets: the Gallup World Poll Cantril ladder
(2020--2022; 135 countries; $N = 300{,}137$), the Gallup satisfaction-with-life item
(2007--2010; 114 countries; $N = 109{,}375$), and the Global Flourishing Study life-today and
satisfaction items (2023; 22 countries;
$N = 177{,}353$ and $177{,}157$). Each fit yields, per
country, the latent (\FVR{}-free) response distribution alongside the
model's predicted reported distribution, with full posterior uncertainty.

Third, empirics: for five inequality statistics I compare cross-country
values and rankings before and after the \FVR{} correction.
I re-run the \GHM{}
country-level regression with raw, predicted, and corrected dispersion
regressors; and I report dominance comparisons
\citep{Jenkins2020-RIW-ordinal} on the corrected distributions.

Fourth, an argument about measures: I argue on a priori grounds
(\Secref{ginicv}) that, just as for the Gini coefficient \citep{springerlink:10.1007/s10902-005-8855-7}, the coefficient
of variation, and the share reporting below~5
are unsuitable for a bounded scale with arbitrary origin:
the first two are mechanically linked to the mean by construction, and the
third saturates toward its bounds as the mean moves away from its fixed
threshold; I accordingly exclude all three from the main analysis of the
mean--dispersion relationship. 

The empirical findings are easily summarized, and are on balance more benign
than the theory might suggest.  \FVR{} inflates measured standard deviation in 133
of 135 Gallup-ladder countries. But the corrections
are nearly uniform in sign, and much of their variation is shared across
countries, so country rankings based on the standard deviation are largely unaffected, 
and the \GHM{} coefficient survives with modest attenuation. The \CF{} ordinal
indices --- the theoretically preferred instruments --- are the most disturbed. \FVR{} alone is thus not sufficient to overturn the existing cross-sectional
correlations. It does, however, further undermine the case for reading any
single scalar of the reported distribution as a measurement of wellbeing
inequality: the candidate statistics are differently contaminated, they
disagree with one another, and those with the most theoretical appeal are
contaminated the most.

In plain terms, the message of the paper is this. A predictable share of
survey respondents simplify the 0--10 scale to just 0, 5, or 10, and because
this behaviour varies with education and possibly other characteristics, it distorts different countries'
reported distributions by different amounts. The distortion is large enough
to move measured inequality levels in nearly every country, but it moves
most countries in the same direction, so orderings and correlations built on
the standard deviation survive largely intact. 
Alternative statistics, designed to be the theoretically safest against scale nonlinearity, turn out to be the most
disturbed by \FVR{}. The contribution is accordingly to improve the measurement and
interpretation of subjective well-being inequality and to clarify what dispersion statistics can and cannot support.
The findings provide reassurance about  additional threats from \FVR{} to broad cross-country
patterns, and caution about the scalar statistics used to summarize them.

The rest of the paper proceeds as follows. \Secref{measures}
reviews the statistics in use and argues for excluding the mean-linked
ones. \Secref{fvr-model} presents the measurement model.
\Secref{fvr-scalar} contains the bound, the analytic bias curves,
and a synthetic falsification of \GHM's inequality-aversion test.
\Secref{empirical} presents the
global estimates, the corrected rankings, the \GHM{} re-estimation, and
the dominance analysis. \Secref{discussion} weighs what survives,
and \Secref{conclusion} concludes.

\section{Statistics of life-satisfaction inequality}\label{sec:measures}

\subsection{The measures in use}

\Swl{} surveys produce, for each country, an empirical distribution
$\{p_k\}_{k=0}^{10}$ over an eleven-point scale.\footnote{0--10 scales are now standard \citep{OECD-2025-guidelines-measuring-SWB-update}, though the World Values Survey still uses a 1--10 scale, and smaller, Likert-style response scales with verbal descriptions for each response option 
  also exist in some surveys.} I focus on five
statistics of that distribution. The within-country {\bfseries standard
deviation} is the measure on which the headline mean--dispersion claims
rest \citep{Goff-Helliwell-Mayraz-EI2018-SWB-inequality}. The
{\bfseries 80:20 difference} is the difference between the survey-weighted
mean response of the top and the bottom quintile of a response
distribution, and is among the measures recommended in practitioner
guidance \citep{WWCW2018}. The status-based indices {\bfseries $I^D_0$
and $I^U_0$} of \citet{CowellFlachaire2017} arose from an 
ordinal-index lineage 
\citet{AllisonFoster2004, AbulNagaYalcin2008, Apouey2007}; in these,  each respondent is scored by her
upward- or downward-looking \emph{status} --- e.g., the share of the
population responding no higher than she does --- rather than by the
numeric label.
The {\bfseries variation ratio} is $1-\max_k p_k$, i.e., the share of
responses away from the modal category. This is the purely ordinal measure
that \GHM{} themselves deploy in defence of their result
\citep{Goff-Helliwell-Mayraz-EI2018-SWB-inequality}. Each of the five is
computed below on synthetic and on empirical data and compared with its
\FVR{}-corrected value. Dominance criteria for ordinal distributions,
applied in \Secref{empirical}, are supplied
by \citet{Jenkins2020-RIW-ordinal} and \citet{GravelMagdalouMoyes2020}.

Other statistics have been applied elsewhere in the context of life satisfaction inequality: for instance, \citet{Stevenson-Wolfers-NBER2008-happiness-inequality-United-States} use the Gini coefficient; \citet{WWCW2018} suggests 
the coefficient of variation, threshold shares, and the Shannon index;
and \citet{DelheyKohler2011} proposes modified standard deviations to correct for structural dependence on the mean. Three of these are excluded from
the main analysis for the reasons developed next.

\subsection{Why the Gini, CV, and threshold share are excluded}\label{sec:ginicv}

The Gini coefficient and the coefficient of variation are
\emph{relative} inequality measures: both are ratios with the mean in the
denominator, and both are therefore invariant to a rescaling of all
values by a common factor. That invariance is their point. For a
ratio-scale quantity such as income --- unbounded above, with a
meaningful zero --- scale invariance is exactly the right requirement: a
currency conversion should not change measured inequality.

A 0--10 response scale has neither property. Its origin is arbitrary: the
same survey item could have been printed 1--11, and nothing in the
responses distinguishes the two labellings. A statistic that is invariant
to multiplying all responses by a constant, but \emph{not} to adding one,
imports the geometry of a ratio scale onto data that possess no zero. The
practical consequence is mechanical: with the mean in the denominator and
the support bounded, cross-country variation in the mean produces
cross-country variation in the Gini and the CV even when absolute
dispersion is identical, and the induced association with the mean is
negative by construction. A regression of mean \SWL{} on the Gini or CV
of \SWL{} therefore has its sign partly built in before any behaviour
enters. The standard deviation is not immune to boundedness either ---
its maximum on a bounded support is itself a function of the mean, the
observation behind the corrected statistic of \citet{DelheyKohler2011}
and the class of normalized indices axiomatized by
\citet{Permanyer-Seth-Yalonetzy-HE2025-inequality-bounded-variables}
--- but the SD at least does not divide by the mean, and it is the
statistic on which the headline empirical claims rest.

A third statistic, the share reporting below~5, is hardly even a measure of inequality when used across populations with varying means. It is the response CDF evaluated at a single, fixed
point: as a country's mean moves away from that threshold in either
direction, the share below it necessarily approaches~0 or~1, whatever the
spread of the remaining mass. This is more than the generic bounded-scale
censoring that afflicts every statistic on this scale
(\Secref{bias-curves-real}): near either tail the statistic is
set almost entirely by location, leaving dispersion almost nothing left
to determine.

Accordingly, these measures are excluded from the main analysis, although it is reproduced for them in \Appref{excludedmetrics}.

\section{Focal-value rounding and a model to correct for it}\label{sec:fvr-model}

\subsection{The phenomenon}

\citet{Barrington-Leigh-JPubEcon2024-focal-values} documents, in
individual-level data from Canada, the United Kingdom, Australia, and the United States,
that responses to 0--10 \SWL{} items concentrate at $\fvset$; that this \FVR{}
reflects a respondent-level propensity to simplify the scale rather than
a property of the underlying wellbeing distribution; and that this
propensity is strongly predicted by education. Reported national
distributions elsewhere show the same spikes
\citep{Barrington-Leigh-DRAFT2026-international-happiness-rankings},
including in some of the  World Values Survey histograms reproduced by
\citet{Grimes-Jenkins-Tranquilli-JoHS2023-SWB-inequality}, and the
estimates below imply that the behaviour is ubiquitous in global survey
data. A respondent who rounds an interior assessment to a focal value
produces a report that no relabelling of the scale can restore: \FVR{}
violates ordinality, the assumption that the ordinal-inequality
literature retains.

\subsection{An adaptation of the \citet{Barrington-Leigh-JPubEcon2024-focal-values} model}

The model in \citet{Barrington-Leigh-JPubEcon2024-focal-values} treats
each respondent as either using the full scale or restricting to
$\fvset$, with a single estimated propensity.\footnote{In the World Values Survey, the focal values are \{1,5,10\}.} The adaptation used here
--- which has not been published elsewhere --- replaces the single
propensity with three: for each focal value $f \in \fvset$, respondent
$i$ retains her full-scale response in the neighbourhood of $f$ with
probability $p^{H,f}_i$, and otherwise rounds responses in that
neighbourhood to $f$. I refer to it as the \emph{reallocation} model.
The symbols introduced below and used throughout are collected in
\Appref{notation}.

\paragraph{Latent layer.}
Respondent $i$ in country $c$ has a latent \SWL{} score
\begin{equation}
  \label{eq:latentnormal}
  \eta_{S,i} = \mu_{S,i} + \varepsilon_{S,i},
  \qquad \mu_{S,i} \equiv X_{S,i}^\top \beta_{S,c},
  \qquad \varepsilon_{S,i} \sim \mathcal N\!\big(0, \sigma_{c}^2\big),
\end{equation}
with $X_S$ containing a constant, log household income, an education
indicator, and gender.
An ordered probit with fixed cardinal cutpoints
$c_k = k - 5.5$ for $k=1,\dots,10$, together with $c_0 = -\infty$ and $c_{11} = +\infty$, maps $\eta_S$ to a full-scale probability mass function ({\bf PMF})
$\boldsymbol\pi = (\pi_0,\dots,\pi_{10})$, with
$\pi_k = \Phi\!\big((c_{k+1}-\mu_{S,i})/\sigma_c\big) - \Phi\!\big((c_k-\mu_{S,i})/\sigma_c\big)$
for $k=0,\dots,10$ --- the Gaussian mass of the latent score falling between
consecutive cutpoints. The scale $\sigma_c$ is the latent dispersion this paper
sets out to measure, so, unlike the textbook ordered probit that normalizes it to
one, it is kept free and country-specific and enters here standardizing the
cutpoint distances. The integer report is thus the interval-censored image of
$\eta_S$ rather than $\eta_S$ itself.

\paragraph{Rounding layer.}
With neighbourhoods $\mathcal N(0) = \{0,1,2\}$,
$\mathcal N(5) = \{3,\dots,7\}$, $\mathcal N(10) = \{8,9,10\}$ and
$p^{H,f}_i = \mathrm{logit}^{-1}(X_{N,i}^\top \beta_{N,f,c})$, the
reported PMF is
\begin{align}
  P(0)  &= \pi_0 + (1-p^{H,0})(\pi_1 + \pi_2), &
  P(j)  &= p^{H,0}\,\pi_j, \; j\in\{1,2\},
  \nonumber\\
  P(5)  &= \pi_5 + (1-p^{H,5})(\pi_3+\pi_4+\pi_6+\pi_7), &
  P(j)  &= p^{H,5}\,\pi_j, \; j\in\{3,4,6,7\},
  \nonumber\\
  P(10) &= \pi_{10} + (1-p^{H,10})(\pi_8 + \pi_9), &
  P(j)  &= p^{H,10}\,\pi_j, \; j\in\{8,9\},
  \label{eq:reallocation}
\end{align}
which preserves total mass. $X_N$ contains a constant and an education
indicator.
Estimation is by Hamiltonian Monte Carlo (NUTS) in PyMC.

\paragraph{Partial-pooling layer.}
The latent coefficients $\beta_{S,c}$, the focal
propensities $\beta_{N,f,c}$, and the latent scale $\sigma_c$ are country-specific but estimated together through partial pooling.
For the question at hand the consequential parameter is the latent scale: $\log\sigma_c$ is drawn from a common hyperdistribution, so each country's latent dispersion is informed by, but not pinned to, the global pattern.
In simulation-based calibration, the specification reproduces
true cross-country $\sigma_c$ variation when it is present and finds none when it
is absent (\Appref{robustness}, \Figref{sigma-recovery}).

\subsection{Identification}\label{sec:identification}

In a traditional ordered logit or probit the cutpoints are free parameters, estimated alongside the latent index. This is also true of the
\citet{Barrington-Leigh-JPubEcon2024-focal-values} \FVR{} mixture model, and a discussion there treats the issue of \FVR{} identification more generally. 
In the reallocation model, free cutpoints exacerbate this challenge: a
cutpoint bordering a focal value may not be separately identifiable from the
rounding propensity toward that value. Specifically, as the top cutpoint
shifts outward, the latent mass at~10 shrinks, but a compensating decrease
in the constant of $\beta_{N}$ raises the rounding propensity, leaving the
reported distribution --- and hence the likelihood --- unchanged along the
ridge.

Pinning the cutpoints to the cardinal positions $c_k = k - 5.5$ removes the likelihood ridge. It fixes the latent metric to the units of the response scale (as in OLS), lets $\sigma_c$ and the focal propensities converge separately, and identifies a constant in $X_S$.
Thus, the FVR{} correction in this paper is identified under
cardinal, equally spaced cutpoints. While addressing the ordinality violation of \FVR{}, the conclusions
drawn below about specifically-ordinal indices are conditional on
that identifying restriction.

\subsection{The objects of analysis}

Estimating the model leaves us with
three PMFs over $\{0,\dots,10\}$: the
raw empirical PMF $\mathbf P^{\text{obs}}_c$; the posterior predicted PMF
$\mathbf{\hat P}^{\text{pred}}_c$, the model's expectation of the
reported data and the natural baseline for isolating the correction; and
the latent PMF $\boldsymbol{\hat\pi}_c$, the counterfactual with
$p^{H,f} \equiv 1$, i.e.\ what would be reported if every respondent used
the full scale. For any statistic $T(\cdot)$ I compare
$T(\mathbf{\hat P}^{\text{pred}}_c)$ with $T(\boldsymbol{\hat\pi}_c)$
across countries. Because both arise from the same posterior, this
comparison isolates the \FVR{} correction, and evaluating both
within each posterior draw gives exact credible intervals for any
cross-country statistic of the pair.

The latent PMF is a model-based object, and it is worth being explicit
about what in it comes from where. What the data identify is the reported
distribution --- the excess mass at each focal value over the smooth
ordered-probit shape, and how that excess varies with education within and
across countries. What the model supplies is equally spaced cutpoints
(\Secref{identification}) and a process in which each respondent either
keeps her full-scale answer or rounds it to the nearest focal value.
$\boldsymbol{\hat\pi}_c$ inherits both, and is an estimate conditional on
that reporting model rather than a model-free fact (\Secref{discussion}).

\section{Theory: How large can the damage be?}\label{sec:fvr-scalar}

This section asks how large the \FVR{} distortion to a dispersion
statistic might be, and it does so without any survey data. I proceed in
two steps. First I derive an analytic bound on the standard-deviation
bias that holds for any latent distribution. Then, because the realized
bias depends on quantities no bound can pin down, I turn to a synthetic
simulation of cross-country data --- a data-generating process with
\emph{no} genuine wellbeing-dispersion channel, specified in full in
\Appref{synthetic-dgp} --- to show how \FVR{} alone can
contaminate cross-country comparisons.
Empirical, survey-based estimates follow later, in \Secref{empirical}.

The role of these exercises is to establish credibility, not estimates, and
each answers one question. The bound says how large the distortion
\emph{could} be. The bias curves say why its size and even its sign depend
on where a country's mean sits relative to the focal values --- which is
what makes the distortion impossible to purge with a linear control. The
synthetic falsification says that a validation test intended to detect
welfare-relevant inequality can be passed in a world containing none.
None of these simulations enters the empirical correction itself, which
depends only on the model of \Secref{fvr-model} and the survey data.

\subsection{A sharp bound on the standard-deviation bias}

The following is the dispersion analogue of the mean-bias bound in
\citet[Proposition~1]{Barrington-Leigh-JPubEcon2024-focal-values}. As
there, the relevant comparison is between what respondents actually
report and what they would report on the full eleven-point scale.

\begin{proposition}[Maximum SD bias due to \FVR]\label{prop:sdbound}
Suppose a fraction $\lambda\in[0,1]$ of respondents report the nearest
element of $\fvset$ (with thresholds at $2.5$ and $7.5$) instead of the
integer they would otherwise give on the full $0$--$10$ scale. Let
$\sdtrue$ be the standard deviation of full-scale responses and $\sdobs$
that of the actual, partially rounded responses. Then
\begin{equation}
  |\sdobs - \sdtrue| \;\leq\; 2\sqrt{\lambda},
  \label{eq:max-sd-bias}
\end{equation}
and the bound is sharp for every $\lambda\in[0,1]$ --- as an increase
or as a decrease, by the $\lambda$-indexed distributions given in the proof
(\Appref{proof}).
\end{proposition}

\begin{remark}
For \FVR{} rates of $\lambda \approx 0.20$--$0.40$, as estimated in
\citet{Barrington-Leigh-JPubEcon2024-focal-values}, the bound permits a
bias of $\sim$0.9--1.3 points on the 0--10 scale --- roughly half the
cross-country range of measured \SWL{} dispersion
(\Tabref{scale}).
The two $\lambda$-indexed families of distributions that attain the bound
(\Appref{proof}) show that \FVR{} can
either inflate or compress the measured SD, depending on where the
population sits relative to the rounding thresholds. The bound is
permissive, not predictive; whether the realized bias approaches it, and
which sign it takes, are questions for the structural estimates of
\Secref{empirical}.
\end{remark}

\subsection{Every dispersion statistic is contaminated by the latent mean}\label{sec:bias-curves}
  
\begin{figure}[p]
\centering
\includegraphics[width=1\linewidth]{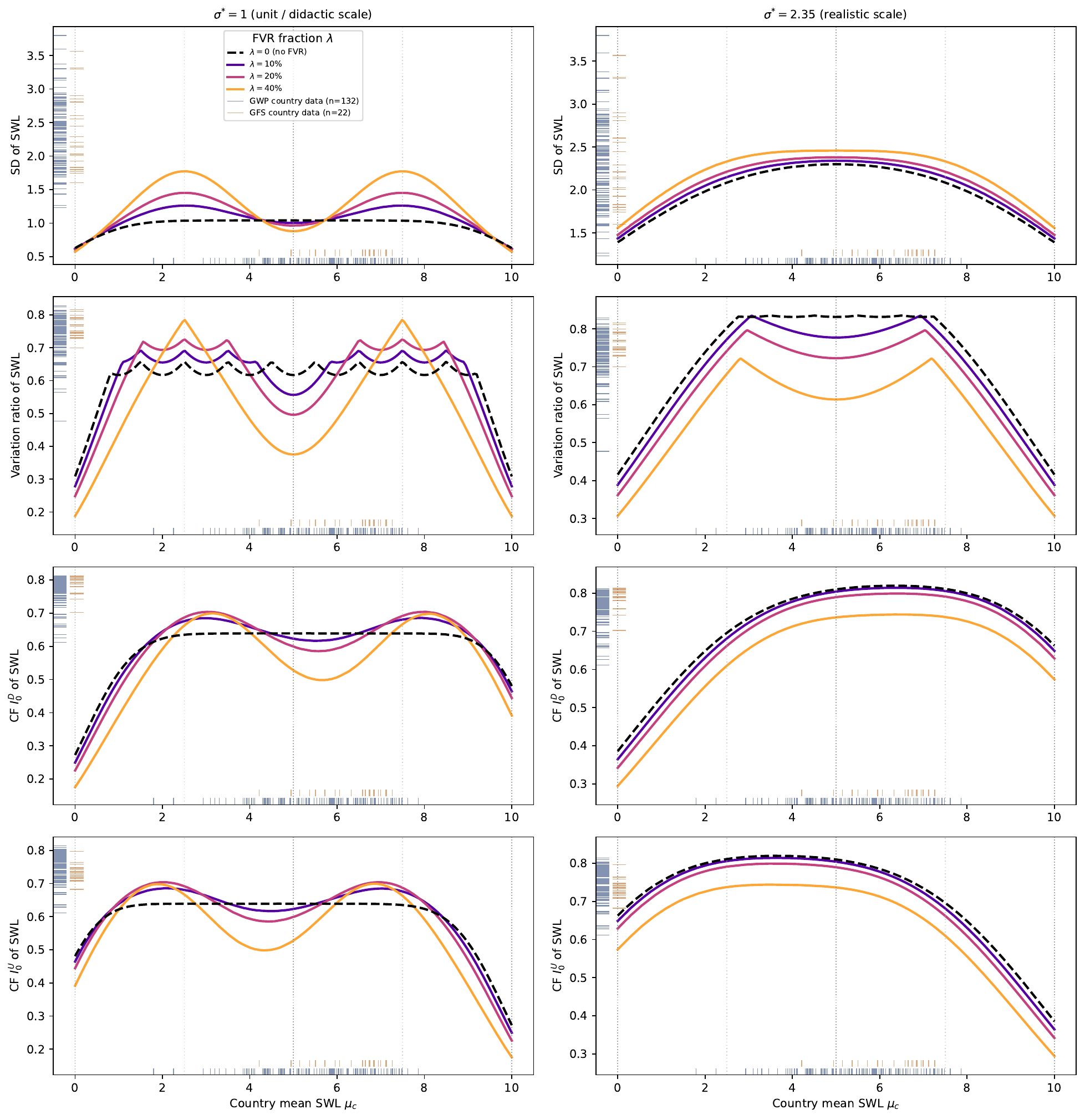}

\caption[Analytic FVR bias curves for four dispersion statistics]{\textbf{Analytic \FVR{} bias curves for four dispersion statistics.} Analytic \FVR{} bias in four dispersion statistics as a function of the
sample mean $\mu$, one curve per focal-responding share
$\lambda\in\{0,0.1,0.2,0.4\}$; there is no sampling noise
(\Appref{dgp-curves}). \textbf{Rows:} standard deviation; the ordinal
variation ratio ($1-\max_k p_k$, the share of responses not at the modal
category); the \CF{} downward-looking status index $I^D_0$; and the \CF{}
upward-looking index $I^U_0$. \textbf{Columns:} the  unit scale
$\sigma^{*}=1$ (left) and the realistic scale
$\sigma^{*}=2.35$ (right) --- the latter the empirical 
mean
within-country \SWL{} standard deviation across the GWP and GFS countries. In each
panel the $\lambda{=}0$ curve is the bounded-scale (censoring) baseline; its
vertical distance to the $\lambda{>}0$ curves is the \FVR{} bias. Marginal rugs
mark the real GWP$+$GFS country means (horizontal axis) and country dispersion
values (vertical axis). The $\lambda$ legend (top-left panel) is common to all
eight panels.}
\label{fig:bias-curves}
\end{figure}

\Figref{bias-curves} evaluates each dispersion statistic\footnote{Two
\href{https://alum.mit.edu/www/cpbl/publications/SWBinequality/}{online
interactive simulations} demonstrate the effects of these measurement problems in
shaping SWB reports and in inducing a mechanical relationship between mean and
dispersion metrics.} on
the exact reported distribution implied by a latent $\mathcal N(\mu,\sigma^{*})$
--- clipped and rounded to the $0$--$10$ integers, then with a fraction $\lambda$
of mass moved to the nearest focal value --- with no sampling noise, and traces it
against the mean $\mu$, one curve per focal-responding share $\lambda$.
The $\lambda{=}0$ curve is the pure bounded-scale
(censoring) baseline; its vertical gap to the $\lambda{>}0$ curves is the \FVR{}
bias.\footnote{Even this baseline is not exactly the latent SD: discretizing a
continuous latent to integers adds Sheppard's correction, so at
$\sigma^{*}=1$ the uncensored reported standard deviation
averages about $1+\tfrac{1}{\sqrt{12}}\approx1.04$
\citep{Sheppard-PLMS1897-moments-of-discrete-distributions}.} The left column of
the figure fixes the latent width at 
$\sigma^{*}=1$; the right column 
raises it to 2.35, as motivated by the empirical country data to follow.

At $\sigma^{*}=1$, the \FVR{} bias in the standard deviation
(\Figref{bias-curves}, top row) is \emph{non-monotone} in the country
mean. 
When the mean sits near the central
focal value 5, \FVR{}  collapses responses onto a single point, leading to compression of the variance. As the mean rises or falls away from the center to 7.5 or 2.5, the variance is exaggerated by \FVR{} moving responses both to 5 and to an extreme, leading
(for $\lambda=40\%$) 
to a maximum bias of
$0.74$.
Finally,  as the mean approaches either extreme the net effect of \FVR{} is increasingly to collapse the distribution towards that extreme, again diminishing the variance.
Because the bias changes sign across the
scale, no single correlation coefficient summarizes it and no linear control for
the mean can purge it.\footnote{The bias can even be non-monotone in the
\emph{degree} of rounding. To see this, consider a latent response that is all
``7''s or all ``8''s: as the \FVR{} fraction rises from~0 to~1, the standard
deviation first rises and then declines.}

The same mean-dependence contaminates the other three statistics, each with its
own geometry. The ordinal variation ratio and the two \CF{} status indices all
vary nonmonotonically with the mean 
--- with peak \FVR{} biases of $0.25$,
$0.14$, and $0.14$ respectively.
The \CF{} indices' invariance to relabelling does
not buy invariance to \FVR{}, because \FVR{} is not a relabelling: it reshapes the
response distribution.

At $\sigma^{*}=1$, the variation ratio  furnishes a relatively clean separation of \FVR{}
from censoring. That is, between means of 1 and 9, its $\lambda{=}0$ curve is unaffected by censoring. Being a
share of responses away from the modal category, it necessarily saw-tooths as
the mean crosses each half-integer and the mode switches to the next category --- a
discreteness artefact that appears even on an unbounded ordinal scale and has
nothing to do with censoring or \FVR{}.

\subsection{At realistic dispersion, censoring dominates and rounding recedes}\label{sec:bias-curves-real}

The unit scale of the left column somewhat isolates the rounding mechanism, but it is likely not a good description of empirical data. The  within-country standard
deviation of raw (reported) life satisfaction across the surveys treated below has average
$\sigma^{*}\approx2.35$,
and at that dispersion
(\Figref{bias-curves}, right column) the emphasis inverts.

First, the \FVR{} bias \emph{washes out}. Its peak in the standard deviation falls
from $0.74$ at unit dispersion to $0.25$;
the variation ratio's peak falls from $0.25$ to
$0.22$; and the \CF{} indices fall from
$0.14$ ($I^D_0$) and $0.14$
($I^U_0$) to $0.10$ and
$0.10$ respectively. A wider latent
distribution straddles several focal neighbourhoods whatever the mean, smearing
away the mean-on-focal structure that drives the swing at
$\sigma^{*}=1$.

Second, and conversely, the \emph{bounded-scale} bias grows and now reaches every
statistic, ordinal or not. The standard deviation's $\lambda{=}0$ (censoring-only)
range across the full scale widens from $0.41$ to
$0.91$, and the variation ratio's from
$0.35$ to $0.42$ (likewise the
\CF{} indices: $I^D_0$ from $0.37$ to
$0.43$ and $I^U_0$ from
$0.37$ to $0.43$;
\Figref{bias-curves}, right column).
More telling than the
aggregate range: at this wider dispersion the tails reach the $0$/$10$ boundary
from means well inside where empirical country-level data sit --- visibly, in the right-column
variation-ratio panel, the $\lambda{=}0$ curve is now bowed well short of the
endpoints, not only at them. The clean ordinal isolation of
\Secref{bias-curves} is thus a  feature of the unit latent
scale: at the empirically-motivated level of dispersion, every measure, ordinal or
not, acquires a mean-dependent censoring bias.

Taken together, the two columns say that, at the scale country data occupies, the
dominant contaminant of the standard deviation is bounded-scale censoring, with
\FVR{} a smaller perturbation on top of it.\footnote{It should be noted, however, that \FVR{}'s relationship to correlates like education may still induce deeper biases.} 
This anticipates the empirical finding of
\Secref{empirical}, where a substantial bounded-scale structure in the
mean--dispersion association coexists with a more modest \emph{net} \FVR{}
correction. The unit-scale column identifies the rounding mechanism; the realistic
column locates its magnitude relative to censoring.

\subsection{\GHM's robustness defences do not neutralize the contamination}\label{sec:ghm-defences}

The robustness arguments this literature deploys are only partly protective.
\citet{Goff-Helliwell-Mayraz-EI2018-SWB-inequality} defend their welfare interpretation
of the mean--dispersion association with three: an allowance for bounded-scale
reporting, a purely ordinal dispersion measure (the variation ratio, the share of
responses away from the mode), and the observation that the association is stronger
among respondents who profess to care most about inequality. None of the three
neutralizes \FVR{}.

Consider the ordinal-measure defence first. As the variation-ratio row of
\Figref{bias-curves} shows (\Secref{bias-curves}
and~\ref{sec:bias-curves-real}), the reported variation ratio swings with the
country mean by up to $0.25$ at unit dispersion, as
$\lambda$ rises, and at
the realistic $\sigma^{*}=2.35$ the measure acquires a genuine
censoring dependence as well.

The bounded-scale allowance, \GHM's first defence, corrects only for a censoring
channel that is common to all respondents; \FVR{}, by contrast, is a reporting
behaviour that differs across respondents --- concentrated  among the less educated
\citep{Barrington-Leigh-JPubEcon2024-focal-values}. Thus,
a bounded-scale correction leaves the \FVR{} contamination untouched.

\GHM's third defence, that the mean--dispersion association is stronger among
respondents who profess to care most about inequality\footnote{
  In most countries of the World Values Survey, higher education is correlated with higher tolerance of inequality according to the measure used by GHM --- as assumed here. However, this question in the WVS is actually phrased around incentives and redistribution, not about concern for inequality (or poverty). These preferences are generally treated as distinct
  \citep{Benabou-Tirole-QJE2006-just-world,Alesina-LaFerrara-JPubE2005-preferences-redistribution}.}
(\GHM's Test~1),
fares no
better in principle. Assuming only that lower-educated respondents exhibit higher \FVR{} rates and 
that the professed
inequality-aversion attitude \GHM's Test~1 relies on is itself correlated with
education ($\rho_a = -0.5$; see \Appref{dgp-aversion}), 
the same interaction they report arises spuriously, with no genuine inequality channel
in effect.
To illustrate this, a regression on synthetic data in which this attitude has no impact is 
shown in \Tabref{ineqav}. The 
relevant interaction coefficient $\beta_{e\sigma} = -0.38$ ($p < 0.001$)
is the same sign as found by \GHM{}, who interpret it as evidence that the inequality-averse perceive real
inequality. 
In this synthetic scenario, the effect operates \emph{within} countries through education being linked at the individual level both to inequality aversion and to \FVR{}. \Appref{ineqav-viz} unpacks this coefficient decile by decile and shows, via a country-level split, that the artefact is indeed carried within countries rather than between them.

  \renewcommand{\ifLongTable}[1]{}
  \renewcommand{\ifNotLongTable}{\cpblEcho}
  \renewcommand{\longTableFooter}{}
\ctResetDefinitions\input{ineqav-regression-table}
\ifcpblTableUseTransposed\cpblTablesSwitchTransposed\fi
\begin{table}
\centering
\ctStartTabular
\ctFirstHeader
\ctBody
\cpblbottomrule 
\ctEndTabular

\cpblBayesColourLegend

\caption[GHM Test 1 falsification regression]{{\bf GHM Test~1 falsification regression.}  OLS of reported life satisfaction $h_{ij}$ on the country reported SD $\sigma_c^{\rm obs}$, simulated inequality aversion $a$, their interaction (column~2), and log income, in a synthetic cross-section with \emph{no} genuine wellbeing--dispersion channel. Column~(2) adds the aversion${}\times{}$SD interaction: $\beta_{e\sigma}=-0.38$ ($p < 0.001$) carries \GHM's Test-1 sign despite the absence of any real inequality channel --- the within-individual artefact visualised in \Figref{ineqav-test}. $N=225{,}000$ individuals; $R^2=0.118$~(1), $0.119$~(2). Standard errors below estimates.      \label{tab:ineqav} \ctDraftComment{[tab:ineqav] [ineqav-regression-table]} \ctCaption }  

\end{table}
\ctResetDefinitionsClosing

The theory and simulations establish what \emph{could} happen. Whether it
does --- whether realized \FVR{} contamination disrupts rankings or
materially distorts the observed mean--dispersion association --- is an
empirical question, treated next.

\section{Empirical estimates}\label{sec:empirical}

\subsection{Data and estimation}

The primary dataset is the Gallup World Poll Cantril ladder (``life
today''), a repeated cross-section pooled over 2020--2022:
$N = 300{,}137$
respondents in 135 countries.
Three other datasets are treated in the Online Appendix and replicate the primary
results. These are the Gallup World Poll satisfaction-with-life item
(2007--2010; $N = 109{,}375$, 114 countries;
\Appref{gwpswl}), which comes from a different period,
and the Global Flourishing Study life-today and
satisfaction-with-life items, asked of the same respondents in a single
cross-section ($N = 177{,}353$ and $177{,}157$;
22 countries; \Appref{gfs}).

The reallocation model is fitted separately to each dataset.
For each country I extract the posterior
predicted and latent PMFs and compute the statistics of
\Secref{measures} on each. Credible intervals for every
cross-country quantity are computed by re-evaluating each within each of 200
posterior draws.%

\subsection{What the distributions look like}

\begin{figure}
\centering
\includegraphics[width=0.95\linewidth]{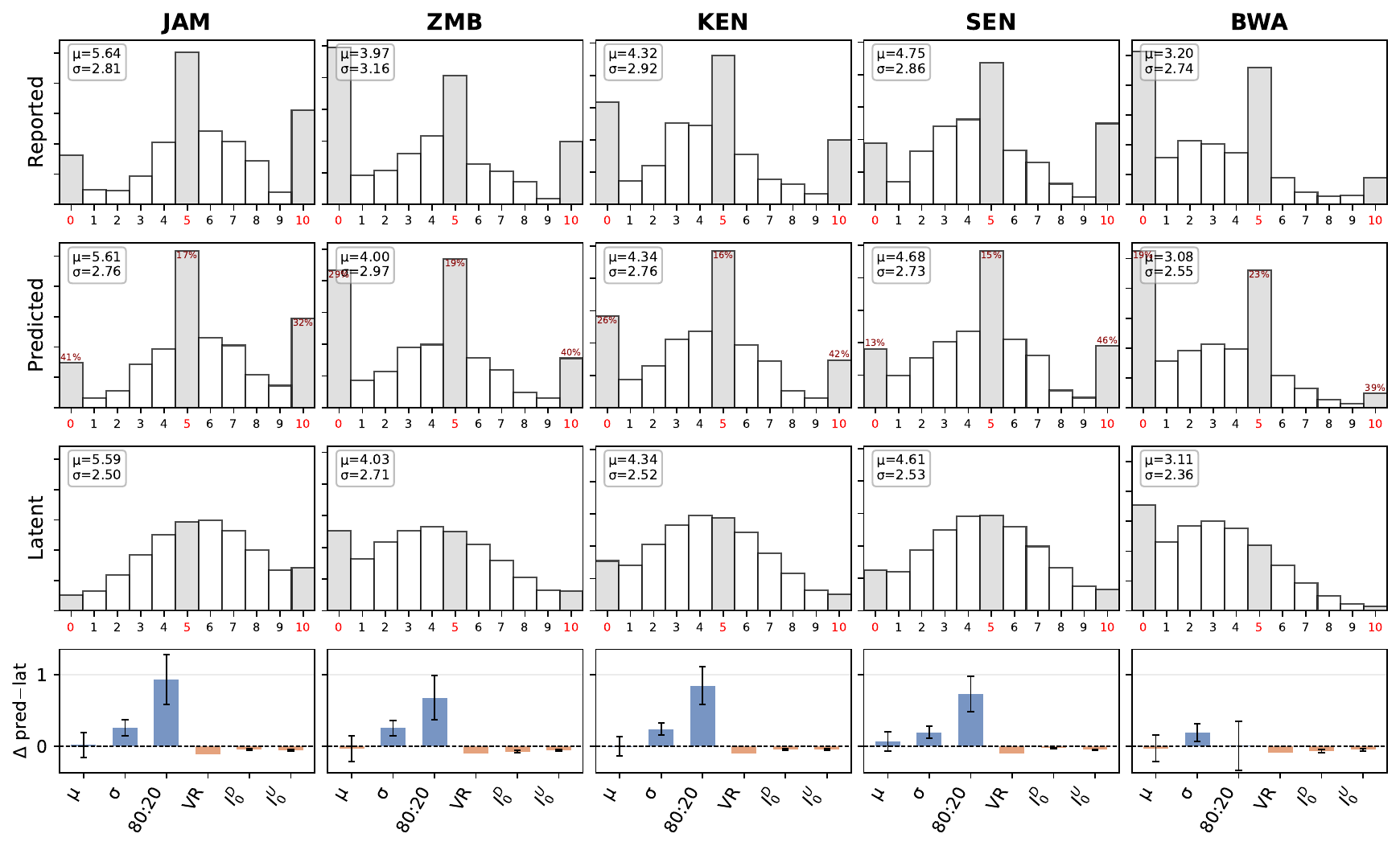}
\caption[Predicted and latent PMFs for countries with the largest corrections]{\textbf{Predicted and latent PMFs for countries with the largest corrections to standard deviation.} Predicted (with \FVR; top) and latent (corrected; middle)
per-country PMFs for the countries with the largest $|\Delta\sd|$, Gallup
Cantril ladder, with per-country differences (predicted $-$ latent) in the
mean and in all five featured statistics --- $\sd$, the 80:20 difference,
the variation ratio, and the \CF{} indices (bottom strip).}
\label{fig:distributions}
\end{figure}

\begin{figure}
\centering
\includegraphics[width=0.95\linewidth]{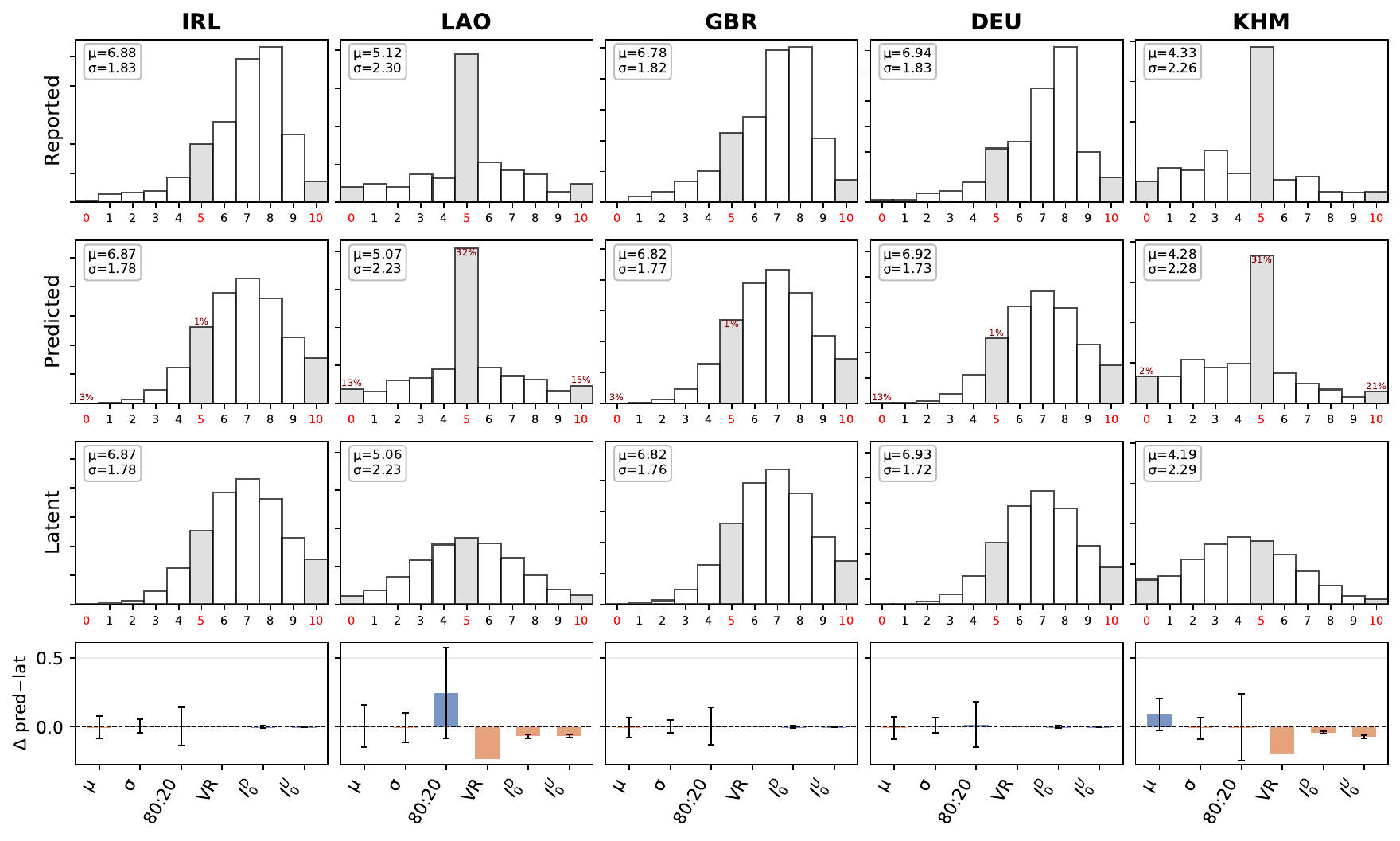}
\caption[Predicted and latent PMFs for countries with the smallest corrections]{\textbf{Predicted and latent PMFs for countries with the smallest corrections to standard deviation.} As \Figref{distributions}, but for the countries with the
\emph{smallest} $|\Delta\sd|$ --- those where \FVR{} barely moves the $\sd$, though that is not necessarily true of the scalar measures shown in the bottom strip.}
\label{fig:distributions-leastchanged}
\end{figure}

\Figref{distributions} shows reported, predicted, and latent PMFs for the countries with the largest  corrections to the standard deviation. The
distributions differ across countries in modes, skewness, and
extreme-category mass in ways that no scalar represents.
\Figref{distributions-leastchanged} shows the opposite extreme ---
the countries where \FVR{} barely moves the dispersion --- for which the
predicted and latent PMFs nearly coincide.
The model allows for two distinct sources of endpoint mass at 0 and 10.
One is boundedness: a latent distribution wide enough to spill past the scale's
endpoints piles up at 0 or 10 under the ordered-probit
censoring, the same
mechanism as in an OLS model. The other is \FVR{}: probability mass from
the neighbourhood of an endpoint is reallocated to the endpoint itself.
Countries with broad latent distributions therefore retain substantial endpoint
mass in $\boldsymbol{\hat\pi}_c$. Such cases are rare ---
\Appref{censoring-exemplars} collects the five countries where the latent
distribution retains the most endpoint mass --- and their rarity is why the
dispersion of the latent PMF tracks the continuous latent scale $\sigma_c$ so
closely (\Secref{sigmac}).

\subsection{The scale and the sign of the correction}\label{sec:scale-sign}

\Tabref{scale} reports, for each statistic: the cross-country
ranges of the predicted and latent versions, the median absolute
correction, the correction as a share of the cross-country range, and
the Spearman correlation between predicted and latent country
rankings.\footnote{The table compares the model's predicted with its
latent rankings because that difference is attributable to the \FVR{}
channel alone. 
}

\begin{table}
\centering
\caption[Scale of the FVR correction, GWP Cantril ladder]{\textbf{Scale of the \FVR{} correction, GWP Cantril ladder.} Scale of the \FVR{} correction in the Gallup World Poll Cantril
ladder ($N=300{,}137$; 135
countries; specification~(A)). Median
$|\Delta|$ is the median across countries of
$|T(\mathbf{\hat P}^{\text{pred}}_c) - T(\boldsymbol{\hat\pi}_c)|$; Rank
$\rho$ is the Spearman correlation between predicted and latent country
rankings.}
\label{tab:scale}
\begin{tabular}{lrrrrr}
\toprule
 & Pred range & Latent range & Median $|\Delta|$ & $\Delta$ / range & Rank $\rho$ \\
\midrule
Mean & {[}1.785, 7.899] & {[}1.794, 7.926] & 0.029 & 0\% & 0.999 \\
SD & {[}1.214, 3.682] & {[}1.166, 3.617] & 0.090 & 4\% & 0.992 \\
80:20 & {[}3.424, 9.713] & {[}3.271, 9.399] & 0.222 & 4\% & 0.988 \\
CF $I^D_0$ & {[}0.622, 0.810] & {[}0.633, 0.834] & 0.019 & 10\% & 0.738 \\
CF $I^U_0$ & {[}0.631, 0.816] & {[}0.648, 0.835] & 0.020 & 11\% & 0.820 \\
Variation ratio & {[}0.496, 0.869] & {[}0.676, 0.880] & 0.075 & 20\% & 0.016 \\
\bottomrule
\end{tabular}
 \end{table}

\begin{figure}
\centering
\includegraphics[width=0.95\linewidth]{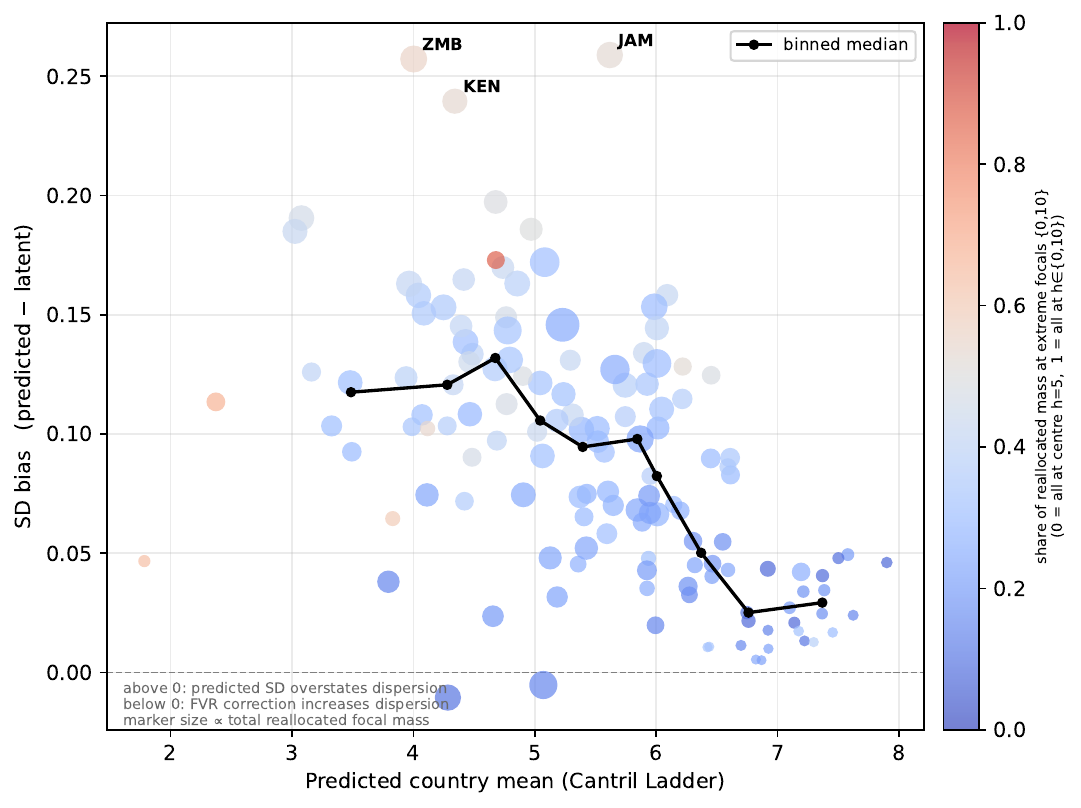}
\caption[Standard-deviation correction versus country mean]{\textbf{Standard-deviation correction versus country mean.} Standard-deviation correction (predicted $-$ latent) versus
predicted country mean, Gallup Cantril ladder. Colour: share of \FVR{}
mass at the extreme focal values $\{0,10\}$ relative to the central focal
5; marker size: total reallocated mass.}
\label{fig:direction}
\end{figure}

Three things stand out in the table:

\paragraph{The correction is material, and almost entirely one-signed.}
\FVR{} inflates the measured standard deviation in
133 of 135
countries (median signed correction
$+0.090$; range
$[-0.01, +0.26]$, the largest
in Jamaica).
That is: in these data, the
analytic possibility of sign-switching is barely realized, because the
effect of rounding toward the extreme focal values --- which stretches the
reported distribution --- dominates that of rounding toward 5 in nearly every country.
\Figref{direction} shows the corrections against the country
mean, coloured by where each country's focal mass sits.

\paragraph{Rankings on the cardinal statistics largely survive.} Because
the corrections are one-signed and partly common, the country ranking on
the standard deviation is nearly preserved: Spearman
$\rho = 0.992$ between predicted and latent (96\%
credible interval
$[0.989,\,0.993]$),\footnote{
  The
96\% level is my chosen convention throughout: credible intervals are
per-draw posterior percentile intervals [2\%ile,98\%ile].
Note also that Spearman point estimates are correlations of the
posterior-mean measures, while the credible intervals are computed from
per-draw correlations, whose mean differs slightly from the point
estimate (a Jensen gap); an interval therefore need not be centred on
the quoted $\rho$.}
with the 80:20 difference behaving similarly
($\rho = 0.988$). The mean is
essentially exact ($\rho = 0.999$; median
$|\Delta| = 0.029$): rounding to 10 raises
it and rounding to 0 lowers it, and the two appear coincidentally to cancel each other to a remarkable degree. Offsetting
biases, in other words, do much of the work that an optimist would have
hoped for --- in levels they do not cancel, but in ranks they largely
do.

\paragraph{The ordinal indices are the most disturbed.} The correlation between uncorrected and corrected \CF{}
downward-looking index is
$\rho = 0.738$ (CI
$[0.688,\,0.764]$)
and that for the upward-looking is $\rho = 0.820$ (CI
$[0.778,\,0.835]$),
on the indices
that the ordinal literature advocates precisely for their robustness
properties.
The mechanism is the one identified in \Secref{fvr-scalar}.
The ordinal indices' merit is that, to make them robust to the critique that response scales should not be interpreted numerically, they are sensitive only to the shape of the PMF, not to any stretching of the scale. However, \FVR{} deforms the cumulative response shares at each category precisely where that cumulative response function changes the fastest --- i.e., at the most-used, focal categories. 

For the same reason, the variation ratio fares worse still: $\rho = 0.016$
(CI $[-0.035,\,0.126]$)
--- statistically indistinguishable from zero, the lowest rank stability of
any of these statistics. This is the real-data counterpart of the synthetic
result of \Secref{ghm-defences}: being ordinal buys the variation
ratio no protection, because \FVR{} routinely relocates a country's modal
response category rather than merely reshaping the mass around it.

The countries where the correction bites hardest, and those where it barely
registers, are listed in \Tabref{topbottom}. The magnitude is
substantial when read as a focal-value response index (\FVRI{}): across the
135 countries the median country relocates onto its
focal values, through \FVR{}, a fraction $0.174$ of its
low-end neighbourhood mass (onto~0), $0.133$ of its
central-neighbourhood mass (onto~5), and $0.139$ (onto~10)
--- where each fraction is the excess focal mass (predicted minus latent) as a
share of that neighbourhood's predicted mass --- for an overall relocated focal
mass of $0.137$ in the median country. Across
countries this overall relocated mass is largest in
Tajikistan.

\begin{table}
\centering
\caption[Largest and smallest SD corrections by country]{\textbf{Largest and smallest standard-deviation corrections by country.} Countries with the largest and smallest standard-deviation corrections
(signed $\Delta\sd = $ predicted $-$ latent), top and bottom of the
135 Gallup Cantril-ladder countries, with the
per-focal-value \FVRI{} (the excess focal mass relocated by \FVR{}, predicted
minus latent, as a share of each neighbourhood's predicted mass); note that
because each \FVRI{} is a share of a possibly tiny neighbourhood mass, a large
percentage can coexist with negligible absolute reallocation --- e.g.,
\FVRI$_0$ in high-mean countries with little low-end mass. Specification
(A).}
\label{tab:topbottom}
\begin{tabular}{lrrrrrr}
\toprule
 & Pred SD & Latent SD & $\Delta$SD & FVRI$_0$ & FVRI$_5$ & FVRI$_{10}$ \\
Country &  &  &  &  &  &  \\
\midrule
Jamaica & 2.757 & 2.498 & +0.259 & 41\% & 17\% & 32\% \\
Zambia & 2.962 & 2.704 & +0.257 & 29\% & 19\% & 40\% \\
Kenya & 2.762 & 2.522 & +0.240 & 26\% & 16\% & 42\% \\
Senegal & 2.728 & 2.530 & +0.197 & 13\% & 15\% & 46\% \\
Botswana & 2.543 & 2.352 & +0.191 & 19\% & 23\% & 39\% \\
Cote d'Ivoire & 2.631 & 2.445 & +0.186 & 19\% & 13\% & 35\% \\
Zimbabwe & 2.389 & 2.204 & +0.185 & 17\% & 22\% & 45\% \\
Mauritania & 3.099 & 2.926 & +0.173 & 8\% & 2\% & 32\% \\
France & 1.678 & 1.667 & +0.011 & 8\% & 2\% & 1\% \\
Spain & 1.713 & 1.702 & +0.011 & 12\% & 1\% & 0\% \\
Italy & 1.649 & 1.639 & +0.011 & 14\% & 1\% & 0\% \\
Germany & 1.730 & 1.720 & +0.010 & 13\% & 1\% & 0\% \\
United Kingdom & 1.765 & 1.760 & +0.005 & 3\% & 1\% & 0\% \\
Ireland & 1.783 & 1.778 & +0.005 & 3\% & 1\% & 0\% \\
Lao PDR & 2.223 & 2.228 & -0.005 & 13\% & 32\% & 15\% \\
Cambodia & 2.281 & 2.291 & -0.010 & 2\% & 31\% & 21\% \\
\bottomrule
\end{tabular}
 \end{table}

A caveat on magnitudes: the comparison above is between the model's
predicted and latent distributions, which isolates the \FVR{} channel.
The gap between the \emph{raw} reported statistics and the latent ones is
wider, but the additional spread reflects model lack-of-fit 
rather than
\FVR{}; conflating the two would overstate the correction.
How much of the relocated focal mass translates into a net correction ---
and how much cancels within a country --- is shown statistic by statistic
in \Appref{fvri-vs-delta}: the net correction tracks the total \FVRI{}
closely for the dispersion and ordinal statistics, but hardly at all for
the mean, where rounding toward 10 and toward 0 largely offset.

\subsection{The \GHM{} regression, before and after correction}\label{sec:ghm-regression}

\citet{Goff-Helliwell-Mayraz-EI2018-SWB-inequality} regress country-mean
\SWL{} on its standard deviation, conditional on log GDP per capita and
(in some specifications) the income Gini. \Tabref{ghm} re-runs the
country-level regression on the Gallup Cantril ladder with three versions
of the dispersion regressor --- raw reported, model-predicted, and latent
--- weighted by country sample size.

\begin{table}
\centering
\caption[GHM-style regression on raw, predicted, and latent SD]{\textbf{\GHM-style regression on raw, predicted, and latent $\sd$.} \GHM-style country-level regression of mean \SWL{} on $\sd$ of
\SWL{}, log GDP per capita, and income Gini, using raw, predicted, and
latent $\sd$ as alternative regressors. Gallup Cantril ladder,
133 countries; WLS by country sample size.
``Implied $\Delta M$ over $\sigma$ IQR'' rescales each coefficient by the
interquartile range of its own regressor. Standard errors, in parentheses, are from re-estimating the regression within each posterior draw.}
\label{tab:ghm}
\begin{tabular}{l|ccc|ccc|ccc}
\toprule
 & \multicolumn{3}{|c|}{raw} & \multicolumn{3}{c|}{pred} & \multicolumn{3}{c}{latent} \\
 & (1) & (2) & (3) & (4) & (5) & (6) & (7) & (8) & (9) \\
\midrule
$\hat\sigma$ raw & \wrapSigFivePercent{$-$.43} &  & \wrapSigOneThousandth{$-$.90} &  &  &  &  &  &  \\
 & \coefse{0.189} &  & \coefse{0.203} &  &  &  &  &  &  \\
\quad Implied $\Delta M$ over $\sigma$ IQR & \wrapSigFivePercent{$-$.33} &  & \wrapSigOneThousandth{$-$.65} &  &  &  &  &  &  \\
 & \coefse{0.145} &  & \coefse{0.145} &  &  &  &  &  &  \\
$\hat\sigma$ predicted &  &  &  & \wrapSigFivePercent{$-$.44} &  & \wrapSigOneThousandth{$-$.86} &  &  &  \\
 &  &  &  & \coefse{0.194} &  & \coefse{0.206} &  &  &  \\
\quad Implied $\Delta M$ over $\sigma$ IQR &  &  &  & \wrapSigFivePercent{$-$.30} &  & \wrapSigOneThousandth{$-$.54} &  &  &  \\
 &  &  &  & \coefse{0.132} &  & \coefse{0.131} &  &  &  \\
$\hat\sigma$ latent &  &  &  &  &  &  & \wrapSigTenPercent{$-$.41} &  & \wrapSigOneThousandth{$-$.82} \\
 &  &  &  &  &  &  & \coefse{0.206} &  & \coefse{0.218} \\
\quad Implied $\Delta M$ over $\sigma$ IQR &  &  &  &  &  &  & \wrapSigTenPercent{$-$.25} &  & \wrapSigOneThousandth{$-$.47} \\
 &  &  &  &  &  &  & \coefse{0.127} &  & \coefse{0.124} \\
Income Gini &  & .008 & \wrapSigTenPercent{.013} &  & .008 & .012 &  & .008 & .011 \\
 &  & \coefse{0.008} & \coefse{0.008} &  & \coefse{0.008} & \coefse{0.008} &  & \coefse{0.008} & \coefse{0.008} \\
Log GDP PPP & \wrapSigOneThousandth{.72} & \wrapSigOneThousandth{.88} & \wrapSigOneThousandth{.57} & \wrapSigOneThousandth{.73} & \wrapSigOneThousandth{.88} & \wrapSigOneThousandth{.61} & \wrapSigOneThousandth{.75} & \wrapSigOneThousandth{.88} & \wrapSigOneThousandth{.65} \\
 & \coefse{0.082} & \coefse{0.059} & \coefse{0.087} & \coefse{0.081} & \coefse{0.059} & \coefse{0.085} & \coefse{0.079} & \coefse{0.059} & \coefse{0.083} \\
Constant & $-$.49 & \wrapSigOneThousandth{$-$3.3} & 1.64 & $-$.59 & \wrapSigOneThousandth{$-$3.3} & 1.16 & $-$.89 & \wrapSigOneThousandth{$-$3.3} & .66 \\
 & \coefse{1.171} & \coefse{0.703} & \coefse{1.282} & \coefse{1.146} & \coefse{0.703} & \coefse{1.251} & \coefse{1.135} & \coefse{0.703} & \coefse{1.239} \\
N & 133 & 121 & 121 & 133 & 121 & 121 & 133 & 121 & 121 \\
$R^2$ & 0.659 & 0.664 & 0.712 & 0.658 & 0.664 & 0.707 & 0.655 & 0.664 & 0.700 \\
\bottomrule
\end{tabular} {\cpblBayesColourLegend}
\end{table}

With the income-Gini control, the coefficient on the raw $\sd$ is
-0.90
$[-1.33,\,
  -0.48]$, on the predicted $\sd$
-0.86
$[-0.92,\,
  -0.72]$, and on the latent $\sd$
-0.82
$[-0.90,\,
  -0.68]$, where each bracketed
range is a 96\% credible interval.\footnote{For the predicted and latent $\sd$
the interval is the posterior credible interval from re-estimating the
regression within each posterior draw; the raw $\sd$ is a fixed function
of the data, so its interval is the classical WLS confidence
interval.} Without the
Gini control the corresponding values are about half as much, with the latent interval
$[-0.48,\,
  -0.30]$.
The cross-sectional \GHM{} association thus survives the \FVR{}
correction in these data: the latent-$\sd$ coefficient is modestly
attenuated relative to the raw one, retains its sign, and remains
conventionally significant. This is an empirical outcome rather than a
foregone conclusion --- \Secref{fvr-scalar} shows that \FVR{}
contaminates cross-country dispersion comparisons in a mean-dependent way that
could have attenuated, inflated, or reversed the association --- and which of
these occurs could only be established by estimating the correction.

Two qualifications keep the result in context. First, the deeper
fragility of the \GHM{} association documented by
\citet{Grimes-Jenkins-Tranquilli-JoHS2023-SWB-inequality} is logically
prior to mine and is untouched by it: in their replication the SD
coefficient loses significance once country fixed effects are included
and is ``virtually zero'' with wave dummies added, and their
theoretically preferred ordinal indices yield significant associations of
\emph{opposite} signs for upward- versus downward-looking status. A
cross-sectional coefficient that survives a measurement correction but
not a fixed-effects specification, or whose sign depends on the choice
among defensible measures, is not yet a foundation for welfare
conclusions. Second, my correction speaks only to the \FVR{} channel;
the cardinality and boundedness critiques of using the SD at all
\citep{%
  Schroder-Yitzhaki-EER2017, DelheyKohler2011}
stand independently.

This sensitivity to the income-Gini control has no counterpart in
\citet{Goff-Helliwell-Mayraz-EI2018-SWB-inequality}: in their
individual-level regressions the standardized coefficient on the \SWL{}
SD is essentially identical with and without the Gini --- $-0.10$ in
both columns for the Gallup World Poll --- and it is instead the Gini
coefficient that shrinks toward insignificance when both enter, a
pattern they read as \SWL{} dispersion subsuming the information in
income inequality. In my country-level cross-section the dependence
runs the other way.

\subsection{The latent scale $\sigma_c$: dispersion free of both rounding and censoring}\label{sec:sigmac}

Every statistic considered so far, including the latent versions, is a
functional of a PMF on the  $0$--$10$ categories, and therefore still
carries the bounded-scale censoring that \Secref{bias-curves-real}
identifies as the dominant contaminant at realistic dispersion. The model,
however, delivers one dispersion object that is free of \emph{both}
distortions: the latent scale $\sigma_c$ of
equation~(\ref{eq:latentnormal}) --- the dispersion that
\Secref{fvr-model} declares this paper's target, measured before the
scale's boundedness or any rounding intervenes. Across the
135 Gallup Cantril-ladder countries, $\sigma_c$
ranges from 1.07 to 5.58, with
median 2.15. Its rank agreement with the latent-PMF
standard deviation is Spearman
$\rho = 0.985$; the distance of that
correlation from one measures how much undoing the censoring, on top of
\FVR{}, reorders countries.

Re-running the \GHM{} regression of \Secref{ghm-regression} with
$\sigma_c$ as the dispersion regressor gives, with the income-Gini control, a
coefficient of -0.43
$[-0.49,\,
  -0.36]$, and without it
-0.17
$[-0.23,\,
  -0.13]$, the bracketed ranges again
being 96\% posterior credible intervals from per-draw re-estimation. This
is the mean--dispersion
association evaluated in the fully corrected metric. Set against the raw,
predicted, and latent-PMF rows of \Tabref{ghm}, these coefficients
show directly how much of the association survives once both
reporting layers --- rounding and censoring --- are removed from the
dispersion regressor, and they suggest a much weaker relationship.
\citet{Goff-Helliwell-Mayraz-EI2018-SWB-inequality} confronted the same
censoring concern by fitting a symmetric logistic latent distribution
in each country--wave cluster and re-estimating in ``actual'' \SWL{}
space; their SD coefficient shrank by roughly a fifth to two fifths
depending on the survey --- from $-0.10$ to $-0.08$ in the Gallup World
Poll --- which they summarize as at most a third of the reported-space
association being a reporting artifact. Their adjustment addresses
censoring under an assumed symmetric latent shape but leaves the
rounding channel untouched, so the $\sigma_c$ comparison here, which
removes both layers, is the more demanding test.

\subsection{Dominance comparisons on the corrected distributions}\label{sec:dominance}

If no scalar is to be trusted, what comparisons remain?
\citet{Jenkins2020-RIW-ordinal} shows that non-intersection of
generalized Lorenz curves of status is equivalent to a unanimous ranking
by the whole \CF{} index family, and finds rankable pairs  in his
six-country application to reported \SWL{} data. I apply
the same criteria to the corrected distributions: for every pair of
countries I test first-order dominance of the response CDFs and
generalized-Lorenz-of-status dominance, on both the predicted and the
latent PMFs. Of 9{,}045 country pairs,
4{,}773
(53\%) are first-order ranked on the
predicted distributions and 6{,}044
(67\%) on the latent ones, with
0 reversals among pairs ranked in both; the
corresponding GL-of-status counts are
516 and
942; the number of reversals among
pairs GL-ranked in both is 1. That GL-of-status
dominance certifies far fewer pairs than first-order dominance is what the
criterion's strength implies: it demands unanimity over the entire \CF{}
index family, and because mean status levels differ across countries, the
generalized Lorenz curves of status cross easily even where the response
CDFs do not. The correction, by removing
spikes that cause CDFs to cross, makes the distributions \emph{more}
often comparable --- and where dominance holds on the latent
distributions, the comparison is robust to every choice of index that
the dominance theorem covers.

\section{Discussion: what survives measurement?}\label{sec:discussion}

\paragraph{The two claims under test.} On the first claim --- that
cross-country variation in a dispersion statistic of reported \SWL{}
reflects underlying wellbeing inequality --- the answer assembled here is
unfavourable, on three converging grounds. First, the mean-linked or
threshold-saturated statistics (Gini, CV, the below-5 share) confuse two
concepts usually intended to be separated (\Secref{ginicv}). Second, the
ordinal alternatives --- the indices designed to accommodate scale nonlinearity, and
\GHM's own variation-ratio defence --- are the statistics most disturbed by
\FVR{} (\Secref{empirical}). And third, the standard deviation, while
rank-stable under the \FVR{} correction,  moves in response to \FVR{}-correction in nearly every country, on top of the
boundedness and cardinality objections that precede this paper.
On the
second claim --- the welfare interpretation of the mean--dispersion association
--- the \FVR{} correction is exculpatory: the
association is not a \FVR{} artifact. Nevertheless, it remains a
cross-sectional association that does not survive fixed effects and whose
sign is measure-dependent in the ordinal family
\citep{Grimes-Jenkins-Tranquilli-JoHS2023-SWB-inequality}.

\paragraph{Practical implications.}
The results suggest an ordering of the response scale's
measurement problems.
\Secref{fvr-scalar} showed that for narrow response distributions, the distortion of standard deviation due to \FVR{} is stronger than that due to censoring (i.e., endpoint effects) everywhere except near the endpoints. However, for broader response distributions like those observed in  country-level data, distortion is dominated by censoring and is significant across the whole scale. The empirical estimates are consistent with the broader distribution case from the simulation: \FVR{} acts mostly as a uniform upward shift in apparent standard deviation, and this shift is small compared with the range of the effect from censoring.
For the other measures considered, the right-hand, broader-distribution column of \Secref{fvr-scalar} is again helpful for understanding the empirical findings. It shows that \FVR{} has a slightly more important effect, though still remarkably uniform, for the \CF{} indices. It also shows that the endpoint effects are even steeper as a function of the mean. For the variation ratio, both the \FVR{} effects and the endpoint effects vary greatly with the mean,  even for the broader, realistic latent width.

What practical implications does this have? 
First, and above all, the issues of end point effects and nonlinear scaling remains an unsolved problem for scalar descriptions of inequality. None of the dispersion measures reflect the latent standard deviation well, though that framing relies on the model of a latent variable extending beyond the limits of cardinal answers. More generally, the unsolved problem reflects the combination of the end point and scaling issues.
Secondly, the \FVR{} correction appears in the cross-country context to matter most for levels rather than ranks. Where
a dispersion statistic is used cardinally --- for instance, its magnitude compared across
populations or over time, or entered as a regressor --- the reported value
should not be treated as a measurement, since \FVR{} inflates the standard
deviation in nearly every country studied here (median
$|\Delta\sd| = 0.090$ on the Gallup ladder), by
amounts that vary across countries.
Third, rank-based uses of standard deviation appear safer based on the present analyses,
unlike the ordinal indices and the variation ratio. However, and fourth, even for ranks the bias from \FVR{} may be larger and more complex
when making comparisons between
populations that (1) plausibly differ in rounding propensity or (2) have a relatively narrow distribution as compared with the whole-country populations investigated here.
The stronger and highly non-monotonic bias evident (again, \Secref{fvr-scalar}) with a narrow distribution means that less confidence is appropriate in measuring inequality 
using the standard deviation when that standard deviation is low --- for instance, below 2.

\paragraph{Limitations.} A main limitation to this work is that the correction is,
inevitably, model-dependent. The \FVR{} reassignment model assumes equally spaced cutpoints for the latent response.
This provides stronger identifiability but may seem a strange assumption with which to evaluate the ordinal \CF{} indices.
In addition, it is behind the interpretation of non-\FVR{} endpoint bunching (censoring) as  a source of dispersion bias as compared with that of an underlying (latent), normally-distributed distribution. A different model of the reporting function might imply a different ``true'' or latent distribution  --- and therefore latent dispersion.
The cardinal-cutpoint reassignment model captures boundedness,
\FVR{}, country-level unexplained variance, and response shape driven by the
distribution of explanatory factors $X$; it does not accommodate response
scale stretching.

On the other hand, this limitation is unlikely to be severe and is likely only to have led to conservative conclusions in this paper.
Interpreting non-\FVR{} endpoint bunching will always be premised on a choice of postulates about the nature of latent wellbeing antecedent to the reporting function.
By positing a normally distributed latent wellbeing as the ultimate object of interest in measuring inequality, along with fixed cutpoints,
the model used here assumes a somewhat maximal role for censoring.  A free-cutpoint ordinal model, for instance, need not recognize the existence of any censoring at all.
This would increase the relative importance of \FVR{} in contributing to any bias.
By comparing the model's latent distribution to its predicted distribution, rather than the reported distribution, in \Secref{empirical}, this paper's analysis has focused on the absolute (not relative) role of \FVR{}. 
Lastly, if scale stretching occurs in
culture- or education-correlated ways, the present correction also {\em understates}
reporting heterogeneity and therefore possible biases contributing to the dispersion--mean relationship.
The future remedy to these issues is continued independent investigation into the response function through innovative techniques along the lines of anchoring
vignettes, qualitative debriefs, and other test--retest designs.

\section{Conclusion}\label{sec:conclusion}

The contributions of this paper are to (i) identify focal-value rounding
as a measurement problem for \SWL{} inequality that is distinct from
 the ordinal-data critique; (ii) bound its possible effect
on the standard deviation at $2\sqrt\lambda$ and show in
simulation that, with the genuine inequality channel switched off by
assumption, \FVR{} contaminates cross-country dispersion comparisons in a
mean-dependent way and can spuriously reproduce \GHM's inequality-aversion-interaction
test;
(iii) estimate the realized contamination at global scale, applying
a \FVR{} correction to cross-country
distributions, calculated with full posterior uncertainty; and (iv) show what the
correction does to the statistics in use: levels move in nearly every country,
cardinal rankings largely survive, the ordinal indices are the most disturbed,
and the \citeauthor{Goff-Helliwell-Mayraz-EI2018-SWB-inequality}
association survives with modest attenuation while remaining fragile on
grounds others have established.

Together, these findings 
provide some improved confidence in the measurement and interpretation of \SWL{} inequality,
but with sharper cautions. The
reassurance for past work is that the headline cross-country patterns are not artifacts
of focal-value rounding: rankings on the standard deviation, and the \GHM{}
association, survive the correction. The caution is that survival of this
kind is not an endorsement of the scalar measures. The surviving statistics
remain subject to the boundedness and cardinality objections that precede
this paper; the theoretically preferred ordinal alternatives are the most
\FVR-disturbed of all; and a statistic robust to one documented reporting
behaviour need not be robust to the next.

My constructive conclusion is not to suggest or seek a better scalar. It is that the distribution itself --- or a
corrected version of it --- is the right object of analysis. Dominance
comparisons on latent distributions resolve more country pairs than on
reported ones and are robust across the entire index family they cover;
distributional plots carry the skewness and extreme-mass features that
scalars discard; and targeted group comparisons can answer policy questions
that no population-wide index addresses.

\clearpage
\printbibliography

\clearpage
\zlabel{page:onlineappendix}
\setcounter{page}{1}
\renewcommand{\thepage}{S\arabic{page}}
\setcounter{section}{0}
\renewcommand{\thesection}{S\arabic{section}}
\renewcommand{\thetable}{\thesection.\arabic{table}}
\renewcommand{\thefigure}{\thesection.\arabic{figure}}
\renewcommand{\theHsection}{S\arabic{section}}
\renewcommand{\theHtable}{S\arabic{section}.\arabic{table}}
\renewcommand{\theHfigure}{S\arabic{section}.\arabic{figure}}
\makeatletter
\@addtoreset{table}{section}
\@addtoreset{figure}{section}
\makeatother

\begin{center}
  {\LARGE\bfseries Online Appendix:\par}
  \vspace{0.7em}
  {\LARGE Does life-satisfaction inequality measure societal inequality?\par}
  \vspace{0.3em}
  {\large A focal-value-rounding critique\par}
  
    \vspace{1.4em}
    {\large C.P.~Barrington-Leigh\par}
    \vspace{1.4em}
    \latestversionline
  
\end{center}
\vspace{1.5em}

\section{Notation}\label{app:notation}

\begin{tabular}{ll}
\toprule
Symbol & Meaning \\
\midrule
$y_i$ & Reported \SWL{} of respondent $i$, integer $\in\{0,\dots,10\}$. \\
$\boldsymbol\pi_c$ & Latent (full-scale) PMF for country $c$. \\
$\mathbf P^{\text{obs}}_c$ & Raw empirical PMF for country $c$. \\
$\mathbf{\hat P}^{\text{pred}}_c$ & Posterior-mean predicted reported PMF. \\
$\boldsymbol{\hat\pi}_c$ & Posterior-mean latent PMF for country $c$. \\
$F_c$ & Response CDF for country $c$. \\
$s_i$ & \CF{} status, $s_i = F(u_i)$ (downward) or $1-F(u_i^-)$ (upward). \\
$f \in \fvset$ & Focal value; $\lambda$ the focal-responding share. \\
$X_S$ & Latent-layer covariates (constant, $\ln$ HH income, education, gender). \\
$X_N$ & Rounding-layer covariates (constant, education).\\ 
$c_k = k - 5.5$ & Cardinal cutpoints, $k=1,\dots,10$ ($c_0=-\infty$, $c_{11}=+\infty$). \\
$\sigma_c$ & Latent scale of country $c$ (hierarchical). \\
$\sd, \sdobs, \sdpred, \sdtrue$ & SD: generic, reported, predicted, full-scale (latent). \\
$I^D_\alpha, I^U_\alpha$ & \CF{} downward-/upward-looking ordinal indices. \\
\bottomrule
\end{tabular}

\section{Proof of \Propref{sdbound}}\label{app:proof}

Let $T_i^{\rm full}$ be the full-scale integer report of individual $i$
and $T_i$ the actual report; for non-\FVR{} respondents these coincide,
while for \FVR{} respondents $T_i \in \fvset$. Define
$D_i = T_i - T_i^{\rm full}$, so $D_i = 0$ for non-\FVR{} respondents.

For an \FVR{} respondent, $T_i^{\rm full}$ is the nearest integer to her
latent assessment and $T_i$ the nearest focal value. The maximum
discrepancy is 2, occurring near the rounding boundaries: a latent value
just below $2.5$ gives $T^{\rm full}=2$, $T=0$ ($D=-2$); just above $2.5$
gives $T^{\rm full}=3$, $T=5$ ($D=+2$); and analogously near $7.5$. Hence
$|D_i| \le 2$ for all respondents, and
\[
  \sigma_D^2 = \operatorname{Var}(D) \;\le\; E[D^2] \;\le\; 4\lambda ,
\]
so $\sigma_D \le 2\sqrt\lambda$. For any random variables $X$ and $Y$,
the $L^2$ triangle inequality gives
$|\sigma_X - \sigma_Y| \le \sigma_{X-Y}$; applying this with $X = T$ and
$Y = T^{\rm full}$ yields
$|\sdobs - \sdtrue| \le \sigma_D \le 2\sqrt\lambda$. The first inequality
binds only if $\operatorname{corr}(T,T^{\rm full})=1$, i.e.\ only if every
respondent's $(T^{\rm full},T)$ pair lies on one common line; this is
automatic when $\lambda=1$, since any two points are collinear, but for
$\lambda<1$ it constrains where the unrounded remainder can sit.

\emph{Tightness, positive direction (\FVR{} inflates the SD), for any
$\lambda$.} Let a $\lambda/2$ share of respondents have latent assessments
just below $2.5$ ($T^{\rm full}=2$, $T=0$, $D=-2$) and a $\lambda/2$ share
just above $7.5$ ($T^{\rm full}=8$, $T=10$, $D=+2$), both \FVR{}; let the
remaining $1-\lambda$ share sit exactly at the scale midpoint
($T^{\rm full}=5$), non-\FVR{} but reporting $5$ regardless since $5\in\fvset$.
Pairing the two extremal points from \emph{opposite} thresholds, rather than
two adjacent to the same threshold, puts the midpoint exactly on the line
joining them, since $5$ is both the arithmetic mean of $2$ and $8$ and of
their images $0$ and $10$; pairing two points adjacent to the same threshold
instead forces the line through them to cross the diagonal at the
non-integer $T^{\rm full}=2.5$, so no unrounded respondent can join it and
that construction attains the bound only at $\lambda=1$. Then
$\sdtrue = 3\sqrt\lambda$, $\sdobs = 5\sqrt\lambda$, and
$\sdobs - \sdtrue = 2\sqrt\lambda$ for every $\lambda\in[0,1]$.

\emph{Tightness, negative direction (\FVR{} compresses the SD), for any
$\lambda$.} Let a $\lambda/2$ share sit just above $2.5$
($T^{\rm full}=3$, $T=5$, $D=+2$) and a $\lambda/2$ share just below $7.5$
($T^{\rm full}=7$, $T=5$, $D=-2$), both \FVR{}; let the remaining
$1-\lambda$ share again sit at the midpoint ($T^{\rm full}=5=T$). The two
\FVR{} branches already share a common rounded image, $5$, so the midpoint
mass leaves $T\equiv 5$ identically. Then $\sdtrue = 2\sqrt\lambda$,
$\sdobs = 0$, and $\sdtrue - \sdobs = 2\sqrt\lambda$ for every
$\lambda\in[0,1]$. \qed

\section{The synthetic data-generating process}\label{app:synthetic-dgp}
This appendix specifies two objects of different kinds. The analytic bias curves
of Figures~\ref{fig:bias-curves} and~\ref{fig:bias-curves-ginicv} need only a
latent mean, an \FVR{} fraction, and a latent dispersion; their construction is
given in \Appref{dgp-curves}. Everything else here --- the individual
layer, the country cross-section, and the inequality-aversion attitude --- serves
\emph{only} the single simulated cross-section behind the \GHM{} Test~1
falsification: the regression of \Tabref{ineqav} and its decile-by-decile
unpacking in \Figref{ineqav-test}.

\subsection{The analytic bias curves}\label{app:dgp-curves}

The bias curves of \Figref{bias-curves} and~\ref{fig:bias-curves-ginicv}
carry no sampling noise and no country, education, or income structure. For a
given latent mean $\mu$ and dispersion $\sigma^{*}$, the reported probability mass
function is formed directly: a latent $\mathcal N(\mu,\sigma^{*})$ is clipped and
rounded to the $0$--$10$ integers, and then a fraction $\lambda$ of its mass is
moved to the nearest focal value ($\{0,1,2\}\!\to\!0$, $\{3,\dots,7\}\!\to\!5$,
$\{8,9,10\}\!\to\!10$). Each dispersion statistic is evaluated on that exact PMF.
The curves sweep $\mu$ over $[0,10]$ at
$\lambda\in\{0,0.1,0.2,0.4\}$ (up to $40\%$), at the two latent
dispersions $\sigma^{*}=1$ and
$\sigma^{*}=2.35$. The $\lambda{=}0$ curve is the bounded-scale
(censoring) baseline; its range over the full sweep measures the censoring bias.
The realistic value $\sigma^{*}=2.35$ is the empirical pooled mean
within-country \SWL{} standard deviation across the GWP and GFS countries.

\subsection{The individual layer}\label{app:dgp-individual}

A respondent $i$ in country $c$ has a latent wellbeing
\[
  z_i = m_c + \varepsilon_i,
  \qquad \varepsilon_i \sim \mathcal N(0,1),
\]
where $m_c$ is the country mean centre, so the within-country latent standard
deviation is exactly $1$ in every country by construction. The full-scale integer
report is $y_i^{*} = \mathrm{clip}\!\big(\mathrm{round}(z_i),\,0,\,10\big)$.
Respondent $i$ also carries education $x_i \sim \mathcal N(\mu_x,1)$ and log
income $\ln y_i \sim \mathcal N(0,\sigma_y^{2})$, with the income dispersion
$\sigma_y$ a single value fixed across all countries; neither enters the
wellbeing equation above. Education governs the rounding (below); log income has
no effect on wellbeing and is retained only as a control in the \GHM{}
falsification regression of \Secref{fvr-scalar}, mirroring their
specification. Focal-value rounding is then applied with the
same neighbourhoods as the estimation model: with probability
$p_i = \mathrm{logit}^{-1}(\gamma_0 - \gamma_1 x_i)$ respondent $i$ replaces
$y_i^{*}$ by its nearest focal value ($\{0,1,2\}\!\to\!0$,
$\{3,\dots,7\}\!\to\!5$, $\{8,9,10\}\!\to\!10$); otherwise she reports
$y_i^{*}$. With $\gamma_1 > 0$ lower-education respondents round more often. At
$x = 0$ --- the cross-country centre of mean education
(\Appref{dgp-country}) --- the focal share is
$\mathrm{logit}^{-1}(\gamma_0)$ (about $0.27$ at $\gamma_0 = -1$, $0.12$ at
$\gamma_0 = -2$); a country with mean education $\overline{x}_c$ has focal
share $\mathrm{logit}^{-1}(\gamma_0 - \gamma_1 \overline{x}_c)$ at its mean.

\subsection{The country cross-section}\label{app:dgp-country}

For a cross-section of $C$ countries, each country's wellbeing centre and mean education
are
\[
  m_c
    = \mathrm{clip}\!\big(\mu_0 + \sigma^{\mathrm{across}}_{\mathrm{swl}}\,u_c,\ 3,\ 9.5\big),
  \qquad
  \overline{x}_c = \sigma^{\mathrm{across}}_x\,v_c,
\]
where $(u_c,v_c)$ is a standard bivariate normal with correlation
$\rho_{xs} = \mathrm{corr}(\overline x_c,m_c)$ --- the single
cross-country education--wellbeing correlation --- $\mu_0$ is a fixed global mean
(the ``base'' of \Tabref{dgp-params}), and
$\sigma^{\mathrm{across}}_{\mathrm{swl}}$ and $\sigma^{\mathrm{across}}_x$ are the
across-country standard deviations of the two dimensions. The wellbeing centre is
thus shifted and clipped while mean education stays centred at zero. Each
country's respondents are then drawn from the individual layer above with this
country centre $m_c$ and mean education $\mu_x = \overline{x}_c$; education thus
varies both within and across countries, while the latent wellbeing dispersion is
held at $1$ everywhere.

\paragraph{The assumed magnitudes.} The latent (focal-free) within-country
dispersion is exactly $\mathrm{SD}(\varepsilon) = 1$ by construction, identical
across countries up to the mechanical compression of the bounded $0$--$10$ scale
near its endpoints. The \emph{assumed} cross-country variation in true wellbeing
dispersion is therefore zero: nothing --- income, education, or attitude ---
enters the wellbeing equation, so there is no genuine dispersion channel of any
kind. Any cross-country mean--dispersion pattern in the simulation without \FVR{} is
thus a bounded-scale censoring artefact, and any further movement once \FVR{} is
applied is focal-value rounding --- neither is a wellbeing-inequality signal.

\subsection{The inequality-aversion attitude}\label{app:dgp-aversion}

For the \GHM{} Test~1 cross-section each respondent is additionally assigned a simulated
inequality-aversion attitude
$a_i = \rho_a\,\tilde x_i + \sqrt{1-\rho_a^{2}}\,\nu_i$,
$\nu_i \sim \mathcal N(0,1)$, with $\tilde x_i$ standardised education and
$\rho_a = -0.5$ (lower education $\to$ more averse). This correlation is an
ingredient of the simulation, not of \GHM's argument: it is what lets
education-driven \FVR{} impersonate an aversion--dispersion interaction, and its
sign is set so the artefact carries the same sign as the interaction \GHM{}
report. Crucially $a_i$ enters \emph{no} equation of the DGP: it has no causal
effect on wellbeing. The genuine ``inequality-averse people perceive real
inequality'' effect that \GHM's Test~1 is designed to detect is therefore
assumed to be exactly zero, so any nonzero interaction the test returns is, by
construction, a focal-value-rounding artefact and not a signal.

\subsection{Parameter settings}\label{app:dgp-params}

\Tabref{dgp-params} lists the settings behind the one simulated
cross-section, that of the \GHM{} Test~1 falsification
(\Tabref{ineqav} and \Figref{ineqav-test}). The cross-section is
centred at $\mathrm{base}=6.0$ with $C=150$ countries (matching \GHM's
cross-country sample size), the regime in which the focal mechanism spans the
inflation/deflation boundary.

\begin{table}[h]
\centering
\caption[Parameter settings, GHM Test 1 synthetic cross-section]{\textbf{Parameter settings, \GHM{} Test~1 synthetic cross-section.} Parameter settings for the \GHM{} Test~1 synthetic cross-section
(\Figref{ineqav-test}). Nothing enters the wellbeing equation (so true
dispersion is fixed), $\gamma_1 = 1$, individual $\sigma_x = 1$, cross-country
$\sigma^{\mathrm{across}}_x = 0.7$, fixed log-income dispersion $\sigma_y = 0.6$,
and seed $42$ (the inequality-aversion attitude uses a secondary seed $9999$).
``base'' is the cross-country mean of $m_c$; $N_c$ is respondents per country.}
\label{tab:dgp-params}
\begin{tabular}{lcccccc}
\toprule
Figure & $C$ & $N_c$ & base & $\sigma^{\mathrm{across}}_{\mathrm{swl}}$
       & $\gamma_0$ & $\rho_{xs}$ \\
\midrule
\ref{fig:ineqav-test} (inequality aversion)  & 150 & 1500 & 6.0 & 0.5 & $-2$ & $0.65$ \\
\bottomrule
\end{tabular}
\end{table}

\section{Decile-level visualization of the synthetic Test~1 artefact}\label{app:ineqav-viz}

\begin{figure}
\centering
\includegraphics[width=0.72\linewidth]{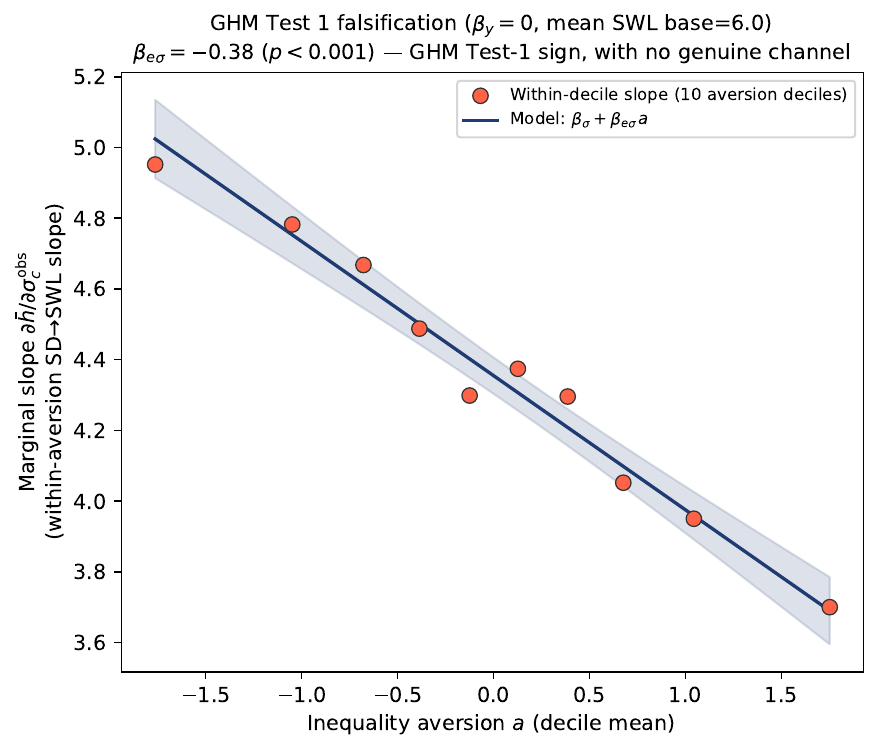}
\caption[Synthetic falsification of GHM's Test 1 interaction]{\textbf{Synthetic falsification of \GHM's Test~1 interaction.} \citeauthor{Goff-Helliwell-Mayraz-EI2018-SWB-inequality}'s Test~1
(inequality-aversion interaction), evaluated in a synthetic cross-section built
with \emph{no} genuine wellbeing-dispersion channel and an
inequality-aversion attitude that has \emph{zero} effect on wellbeing by
construction (\Appref{synthetic-dgp}). Each point is one
inequality-aversion decile; its height is the SD--SWL slope
($\partial\bar h/\partial\sigma_c^{\rm obs}$) estimated \emph{within} that decile.
The slopes fall monotonically as professed aversion rises, and the slope of that
decline is the interaction coefficient $\beta_{e\sigma} = -0.38$ ($p < 0.001$) ---
\GHM's sign --- even though the DGP contains no genuine dispersion channel, so
\FVR{} alone spuriously reproduces the test. (A flat line would mean no interaction; a coarse country-level median
split, by contrast, shows near-equal subgroup slopes because the artefact is
carried within countries by individual education, not by the between-country
slope.)}
\label{fig:ineqav-test}
\end{figure}

\Tabref{ineqav} reports a single pooled interaction coefficient,
$\beta_{e\sigma}$; \Figref{ineqav-test} unpacks that coefficient by estimating
the SD--SWL slope separately within each inequality-aversion decile. Two things
are visible here that the table alone does not show. First, the decline is
monotonic across all ten deciles, not merely consistent with a linear
interaction term. Second, and more informative for locating the mechanism, a
coarse country-level median split on the same synthetic data yields
near-equal subgroup slopes: the artefact is carried \emph{within} countries by
individual education rather than by any between-country composition effect,
confirming the individual-level channel described in \Secref{ghm-defences}
rather than a spurious cross-country correlation.

\section{Bias geometry of the excluded statistics}\label{app:excludedmetrics}

\begin{figure}[h]
\centering
\includegraphics[width=0.85\linewidth]{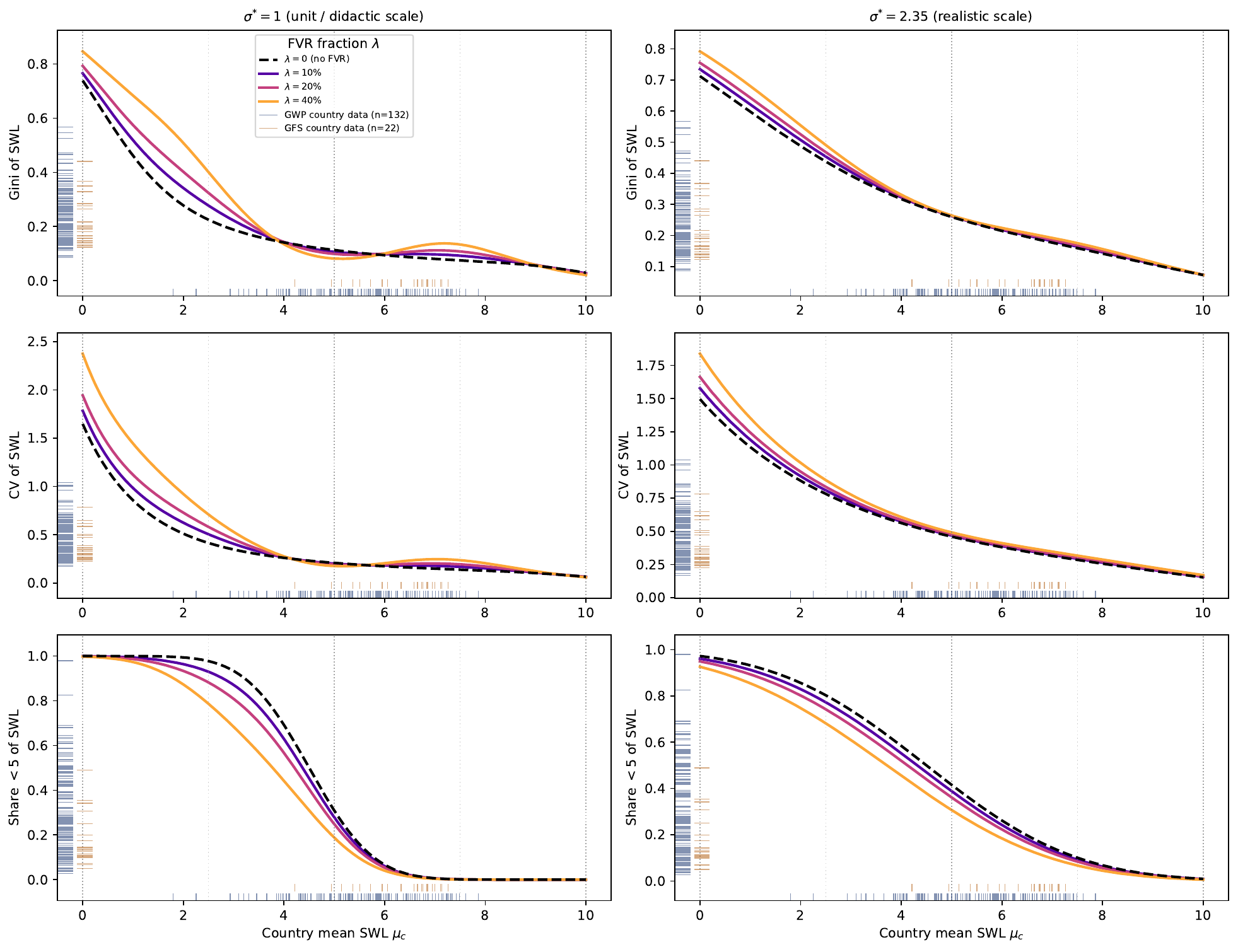}

\caption[FVR bias curves for the excluded statistics]{\textbf{\FVR{} bias curves for the Gini coefficient, the CV, and the below-5 share.} \FVR{} bias in the Gini coefficient, the coefficient of
variation, and the share reporting below~5, as a
function of the country mean $\mu$, one curve per focal-responding share
$\lambda\in\{0,0.1,0.2,0.4\}$, at the  unit scale
$\sigma^{*}=1$ (left) and the realistic scale
$\sigma^{*}=2.35$ (right); construction as in
\Figref{bias-curves}, whose $\lambda$ legend (top-left panel) applies
here too. The Gini and CV divide by the mean; the below-5 share is instead
evaluated at a fixed threshold --- each carries a
strong, mechanical dependence on the mean, unrelated to any change in true dispersion.}
\label{fig:bias-curves-ginicv}
\end{figure}

\Secref{ginicv} excludes the Gini coefficient, the coefficient of
variation, and the share reporting below~5 from the mean--dispersion analysis,
on three related but distinct a priori grounds. For completeness,
\Figref{bias-curves-ginicv} traces the synthetic \FVR{} bias in all
three over the same country-mean sweep as \Figref{bias-curves}, at
both latent dispersions. The Gini and CV divide by the mean; the below-5 share
instead saturates toward~0 or~1 as the mean moves away from its fixed
threshold. Different mechanisms, same outcome, visible in every panel below: a
strong dependence on the mean that has nothing to do with any change in true
dispersion.

\section{Rare cases: latent distributions with strong endpoint mass}\label{app:censoring-exemplars}

\begin{figure}
\centering
\includegraphics[width=0.95\linewidth]{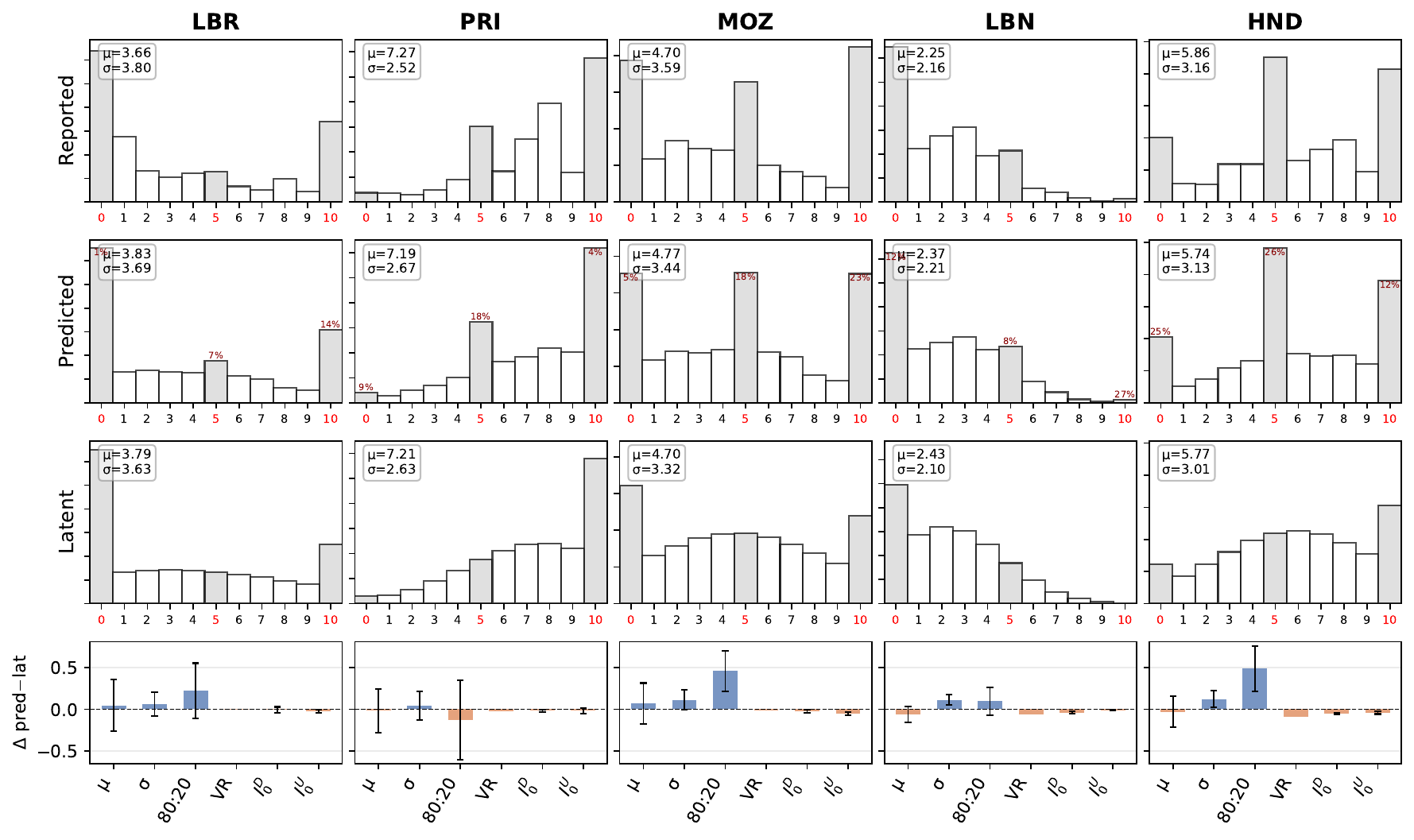}
\caption[Latent censoring exemplars]{\textbf{Rare cases where the latent
distribution retains strong endpoint effects.} The five Gallup
Cantril-ladder countries with the largest latent endpoint mass
$\hat\pi_c(0)+\hat\pi_c(10)$: Liberia (0.45), Puerto Rico (0.31),
Mozambique (0.28), Lebanon (0.25), and Honduras (0.21). Rows show the
reported, predicted, and latent PMFs, with the correction strip below, in
the format of \Figref{distributions}. Even
after the \FVR{} component is removed, these latent distributions are wide
enough that censoring piles mass at the scale endpoints; Puerto Rico is
almost purely a top-end case and Lebanon almost purely bottom-end, while
Liberia, Mozambique, and Honduras show mass at both ends.}
\label{fig:censoring-exemplars}
\end{figure}

The correction leaves most countries' latent distributions with little mass
at the scale endpoints. The exceptions are informative.
\Figref{censoring-exemplars} shows the five Gallup Cantril-ladder countries
--- of the seven candidates with visibly heavy latent tails --- with the
largest latent endpoint mass $\hat\pi_c(0)+\hat\pi_c(10)$. These latent
distributions are wide enough that ordered-probit censoring piles mass at 0
or 10 even after the \FVR{} component is removed: qualitative evidence that
something like censoring is at work in these data. Such cases are rare,
which is also why the dispersion computed from the latent PMF stays close
to the continuous latent scale $\sigma_c$ for almost every country
(\Secref{sigmac}).

\section{How much the corrections cancel: total FVRI versus the net correction}\label{app:fvri-vs-delta}

\begin{figure}
\centering
\includegraphics[width=0.95\linewidth]{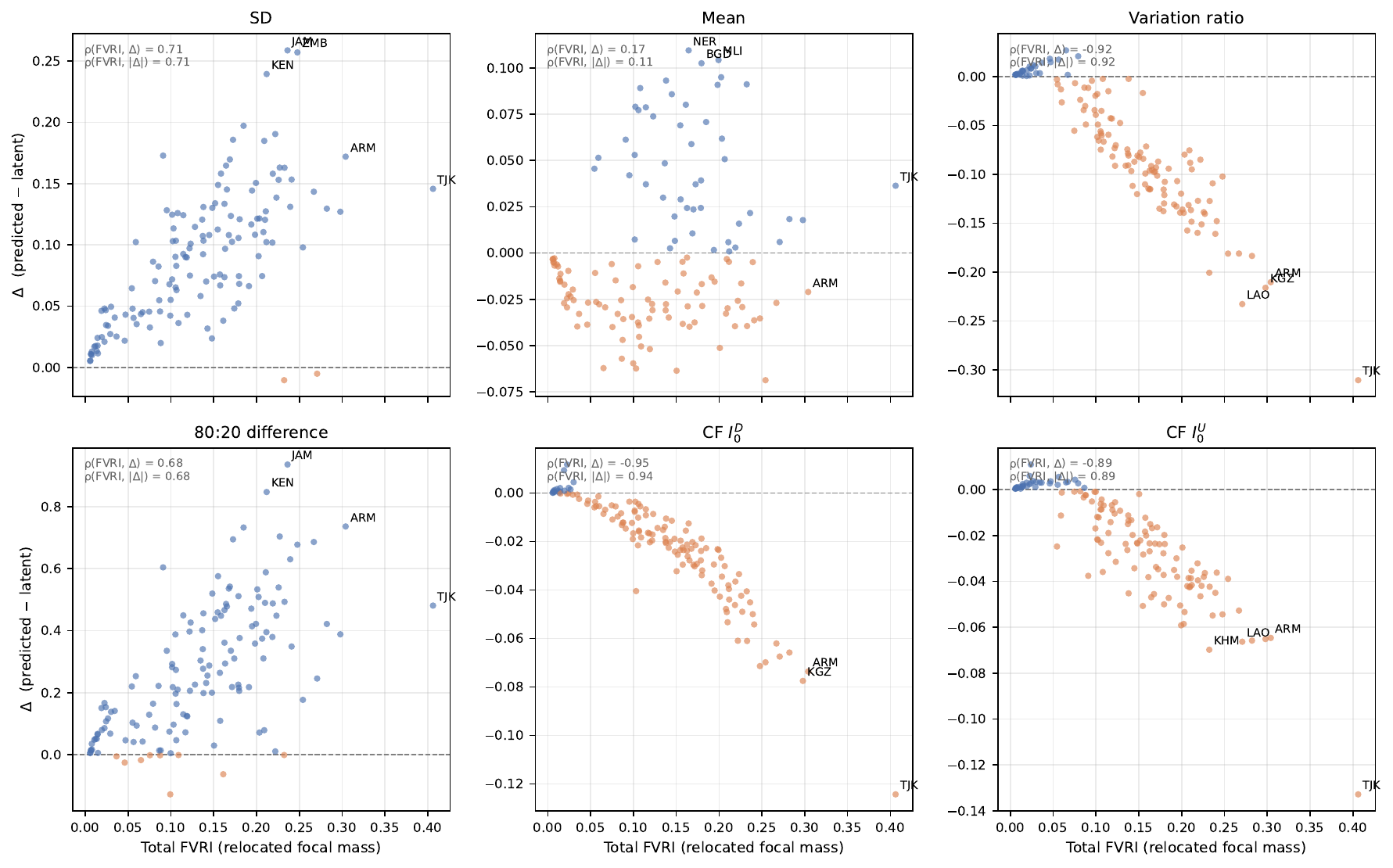}
\caption[Total FVRI versus the correction, by statistic]{\textbf{How much
the \FVR{} corrections cancel.} Each panel plots, per country, the signed
correction $\Delta = $ predicted $-$ latent for one statistic against the
country's overall \FVRI{} (total relocated focal mass), Gallup Cantril
ladder. For the dispersion and ordinal statistics the correction grows
nearly monotonically with the relocated mass (Spearman correlations
annotated), whereas for the mean a large relocated mass yields almost no
net correction: rounding toward 10 and toward 0 largely offset. Extreme
countries are labelled.}
\label{fig:fvri-vs-delta}
\end{figure}

\Figref{fvri-vs-delta} plots, for each statistic, each country's net
correction against its overall relocated focal mass. A statistic could in
principle absorb a large relocated mass with no net change --- the
corrections at the three focal values can cancel --- and the figure shows
where that happens: for the mean, and essentially nowhere else.

\clearpage

\section{Replication on the Gallup satisfaction-with-life item}\label{app:gwpswl}

The Gallup World Poll fields a satisfaction-with-life item alongside the Cantril
ladder, though in a different period and survey mode (2007--2010;
$N = 109{,}375$, 114 countries). I fit the same
reallocation model (specification~(A)) and treat the
result as an independent replication of the primary Cantril-ladder analysis
rather than as an item comparison, since the period and mode differences
confound any direct comparison of the two items.

\begin{table}[h]
\centering
\caption[Scale of the FVR correction, GWP SWL item]{\textbf{Scale of the \FVR{} correction, Gallup satisfaction-with-life item.} Scale of the \FVR{} correction, Gallup satisfaction-with-life item
($N=109{,}375$; 114 countries, 2007--2010).
Columns as in \Tabref{scale}.}
\label{tab:scaleGWPSWL}
\begin{tabular}{lrrrrr}
\toprule
 & Pred range & Latent range & Median $|\Delta|$ & $\Delta$ / range & Rank $\rho$ \\
\midrule
Mean & {[}2.409, 8.403] & {[}2.487, 8.423] & 0.014 & 0\% & 1.000 \\
SD & {[}1.074, 2.817] & {[}1.003, 2.743] & 0.039 & 2\% & 0.984 \\
80:20 & {[}2.897, 8.019] & {[}2.709, 7.750] & 0.081 & 2\% & 0.985 \\
CF $I^D_0$ & {[}0.622, 0.824] & {[}0.627, 0.829] & 0.003 & 2\% & 0.897 \\
CF $I^U_0$ & {[}0.549, 0.818] & {[}0.550, 0.833] & 0.002 & 1\% & 0.954 \\
Variation ratio & {[}0.510, 0.858] & {[}0.615, 0.867] & 0.011 & 3\% & 0.568 \\
\bottomrule
\end{tabular}
 \end{table}

The primary results carry over. The standard-deviation correction is one-signed
--- \FVR{} inflates the reported $\sd$ --- in 103 of
114 countries (median signed
$\Delta\sd = +0.038$, largest $+0.26$ in
Lao PDR), and the country ranking on the standard deviation
survives the correction (Spearman $\rho = 0.984$, 96\% credible
interval $[0.970, 0.987]$). As in the
main analysis the \CF{} ordinal indices are more affected
($I^D_0$: $\rho = 0.897$;
$I^U_0$: $\rho = 0.954$), though the effect on rankings is minimal. The \GHM-style country regression
(111 countries) gives raw/predicted/latent SD
coefficients of -0.20/%
-0.19/-0.20
(latent CI $[-0.28,\,
-0.06]$), so the cross-sectional association
again survives the correction, with no significant attenuation.

\section{Replication on the Global Flourishing Study}\label{app:gfs}

The Global Flourishing Study's first wave asks both the Cantril ``life today'' item
and the satisfaction-with-life item of the same respondents (22 countries; $N = 177{,}353$ and
$177{,}157$ respectively), although the survey question wording somewhat muddles the normal distinction \citep[See Section 2 of ][]{Barrington-Leigh-DRAFT2026-international-happiness-rankings}.
This same-respondent design removes the period and question-wording
confounds that muddy the comparison between the two Gallup datasets.

\begin{table}[h]
\centering
\caption[Scale of the FVR correction, GFS life-today item]{\textbf{Scale of the \FVR{} correction, GFS life-today item.} Scale of the \FVR{} correction, GFS life-today item
($N=177{,}353$; 22 countries). Columns as in
\Tabref{scale}.}
\label{tab:scaleGFSLT}
\begin{tabular}{lrrrrr}
\toprule
 & Pred range & Latent range & Median $|\Delta|$ & $\Delta$ / range & Rank $\rho$ \\
\midrule
Mean & {[}4.304, 7.366] & {[}4.239, 7.367] & 0.014 & 0\% & 0.999 \\
SD & {[}1.509, 3.531] & {[}1.505, 3.481] & 0.049 & 2\% & 0.998 \\
80:20 & {[}4.176, 9.713] & {[}4.166, 9.591] & 0.048 & 1\% & 0.994 \\
CF $I^D_0$ & {[}0.714, 0.819] & {[}0.741, 0.833] & 0.005 & 5\% & 0.784 \\
CF $I^U_0$ & {[}0.688, 0.797] & {[}0.712, 0.824] & 0.004 & 4\% & 0.718 \\
Variation ratio & {[}0.689, 0.833] & {[}0.742, 0.871] & 0.010 & 7\% & 0.036 \\
\bottomrule
\end{tabular}
 \end{table}

\begin{table}[h]
\centering
\caption[Scale of the FVR correction, GFS SWL item]{\textbf{Scale of the \FVR{} correction, GFS satisfaction-with-life item.} Scale of the \FVR{} correction, GFS satisfaction-with-life item
($N=177{,}157$). Columns as in \Tabref{scale}.}
\label{tab:scaleGFSSWL}
\begin{tabular}{lrrrrr}
\toprule
 & Pred range & Latent range & Median $|\Delta|$ & $\Delta$ / range & Rank $\rho$ \\
\midrule
Mean & {[}5.082, 8.029] & {[}5.085, 8.065] & 0.010 & 0\% & 0.999 \\
SD & {[}1.614, 3.720] & {[}1.611, 3.567] & 0.050 & 2\% & 0.997 \\
80:20 & {[}4.412, 10.000] & {[}4.407, 9.777] & 0.069 & 1\% & 1.000 \\
CF $I^D_0$ & {[}0.701, 0.822] & {[}0.757, 0.837] & 0.003 & 2\% & 0.862 \\
CF $I^U_0$ & {[}0.529, 0.783] & {[}0.557, 0.787] & 0.000 & 0\% & 0.879 \\
Variation ratio & {[}0.561, 0.853] & {[}0.592, 0.880] & 0.000 & 0\% & 0.910 \\
\bottomrule
\end{tabular}
 \end{table}

The primary results carry over. Corrections are one-signed in every GFS country
(22 of 22 on the life-today
item; 22 of 22 on the
satisfaction item). SD rankings
survive (GFS LT: $\rho = 0.998$, CI
$[0.991, 0.999]$; GFS SWL:
$\rho = 0.997$), levels move (GFS LT median
$|\Delta\sd| = 0.049$, largest
$+0.21$ in Nigeria), and the \CF{}
indices are most affected (GFS LT $I^D_0$:
$\rho = 0.784$, CI
$[0.670, 0.828]$;
$I^U_0$: $\rho = 0.718$); the rank instability is stronger
in this 22-country sample than in the
135-country Gallup sample, perhaps simply because rank statistics in
small samples are less stable. The \GHM-style regression on GFS
(22 countries) gives raw/predicted/latent SD
coefficients of -1.00/%
-1.05/-0.96
(latent CI $[-1.08,\,
-0.82]$).
The satisfaction-with-life item of the same study behaves the same way: levels
move (median $|\Delta\sd| = 0.050$, largest
$+0.19$ in Kenya), the \CF{} indices
are most affected ($I^D_0$: $\rho = 0.862$; $I^U_0$:
$\rho = 0.879$), and the \GHM-style regression
(22 countries) gives raw/predicted/latent SD coefficients
of -1.48/-1.59/%
-1.58 (latent CI
$[-1.70,\,
-1.43]$), so the cross-sectional association
again survives the correction.

\section{Why the variation ratio is rank-unstable for life-today but not satisfaction-with-life}\label{app:vr-mechanism}

\Secref{scale-sign} reports that the variation ratio's predicted-to-latent
rank correlation on the Gallup Cantril ladder is statistically
indistinguishable from zero ($\rho = 0.016$, CI
$[-0.035,\,0.126]$).
The same collapse recurs on the Global Flourishing Study's life-today item
($\rho = 0.036$, CI
$[-0.076,\,0.189]$;
\Tabref{scaleGFSLT}), yet the satisfaction-with-life item is far more
rank-stable on both surveys --- Gallup $\rho = 0.568$ (CI
$[0.532,\,0.683]$;
\Tabref{scaleGWPSWL}), GFS $\rho = 0.910$ (CI
$[0.809,\,0.945]$;
\Tabref{scaleGFSSWL}). Every other statistic in \Tabref{scale},
\Tabref{scaleGWPSWL}, \Tabref{scaleGFSLT}, and \Tabref{scaleGFSSWL} retains
a rank correlation of at least $0.7$, so the variation ratio's life-today
collapse is a genuine outlier, consistent across two independent surveys,
and one that a defender of \GHM's ordinal-measure argument might
reasonably ask me to explain rather than merely report.

The mechanism is a discontinuity, not a magnitude. The variation ratio is
$1 - \max_k p_k$: a function of only the single largest category share, so
it is indifferent to how a correction reshapes the rest of a country's
distribution but maximally sensitive to whichever shift, however small,
changes \emph{which} category is largest. The predicted and latent PMFs
disagree about the modal category in
73 of
135 Gallup life-today countries and
8 of 22 GFS
life-today countries, against only 35
of 114 Gallup satisfaction-with-life countries and
2 of 22 GFS
satisfaction-with-life countries.

What separates the two items is how close each country's top two
categories already are before any correction is applied. The margin
between a country's largest and second-largest latent-PMF share has a
median of only 0.009 on Gallup
life-today (mean 0.018) and
0.010 on GFS life-today --- a typical country's
top two response categories separated by roughly one percentage point of
the response mass --- against 0.014 and
0.019 respectively for satisfaction-with-life.
A wide margin survives a small correction; a razor-thin one does not. And
the correction that closes those thin life-today margins is not generic:
category~5, the central focal value already implicated in
\Secref{ginicv}'s below-5-share argument, is the modal category of the
\emph{predicted} (rounding-contaminated) distribution in
100 of
135 Gallup life-today countries
($74\%$), even though the latent mode
ranges over 8 distinct categories.
\FVR{} manufactures a spike at 5 that the de-rounded, latent distribution
does not have; landing on a latent distribution that is itself often close
to flat-topped across two or three adjacent categories, that manufactured
spike is frequently enough to relocate the identified mode.
Satisfaction-with-life responses are not similarly exposed: on GFS, for
instance, category $10$ --- the top of the scale,
not the centre --- is the predicted mode in only
$50\%$ of countries, and the corresponding
latent margin (0.019) is wide enough to
survive the correction intact in nearly every case.

The instability is a property of the correction itself, not of the
model's forward simulation. In every dataset, mismatches between the
\emph{raw, reported} modal category and the latent one are at least as
large as predicted-versus-latent mismatches (Gallup life-today:
87 of
135 raw--latent mismatches against
25 raw--predicted mismatches; GFS
life-today: 12 of 22
against 6). Predicted tracks raw closely,
as the model is fit to reproduce the observed distribution; it is
specifically the raw/predicted-to-latent step --- the \FVR{} correction
--- that relocates the mode. That the pattern recurs on both surveys,
despite their different periods, modes, and question wordings, rules out
a survey-specific artefact: what the two life-today items share is the
item itself, a Cantril-ladder framing whose responses spread across the
upper-middle of the 0--10 scale, straddling the central focal value,
rather than clustering as decisively on a single category as
satisfaction-with-life responses do.

None of this weakens the claim in \Secref{ghm-defences}: the Gallup
life-today variation ratio's rank instability is real, replicates on an
independent survey, and now has a concrete mechanism rather than only a
synthetic analogue. But the mechanism is contingent on how decisively a
population's modal response category is separated from its runner-up, not
a generic property of ordinal, mode-based statistics under \FVR{}. Being
ordinal is no defence against \FVR{} for the Gallup life-today item at the
centre of \GHM's claim; it should not, however, be read as evidence that
every ordinal measure is this fragile for every item.

\section{Specification robustness and estimate quality}\label{app:robustness}

The primary estimates use specification~(A):
hierarchical per-country
$\log\sigma_c$, country-varying $\beta_S$ and $\beta_N$, cardinal
cutpoints, and rounding-layer covariates comprising a constant and an
education indicator. A companion specification~\specB,
differing in the rounding-layer covariates and in a looser $\beta_S$ prior,
was run on all four datasets as a robustness check: per-country latent
dispersion agrees across the two specifications at Spearman rank correlation
$0.9998$ and Pearson $0.9999$ across all Gallup countries, with every
discrepancy an order of magnitude inside the posterior standard error.
Convergence: all chains pass divergence and posterior-predictive checks
on all four datasets; diagnostics flag slow mixing (low effective sample
size) for the per-country latent-scale parameters of a cluster of
low-dispersion countries in the primary Gallup dataset, which affects the
precision of those countries' interval statements but --- per the
cross-specification agreement above --- not the paper's cross-country
statistics.

\begin{figure}
\centering
\includegraphics[width=\linewidth]{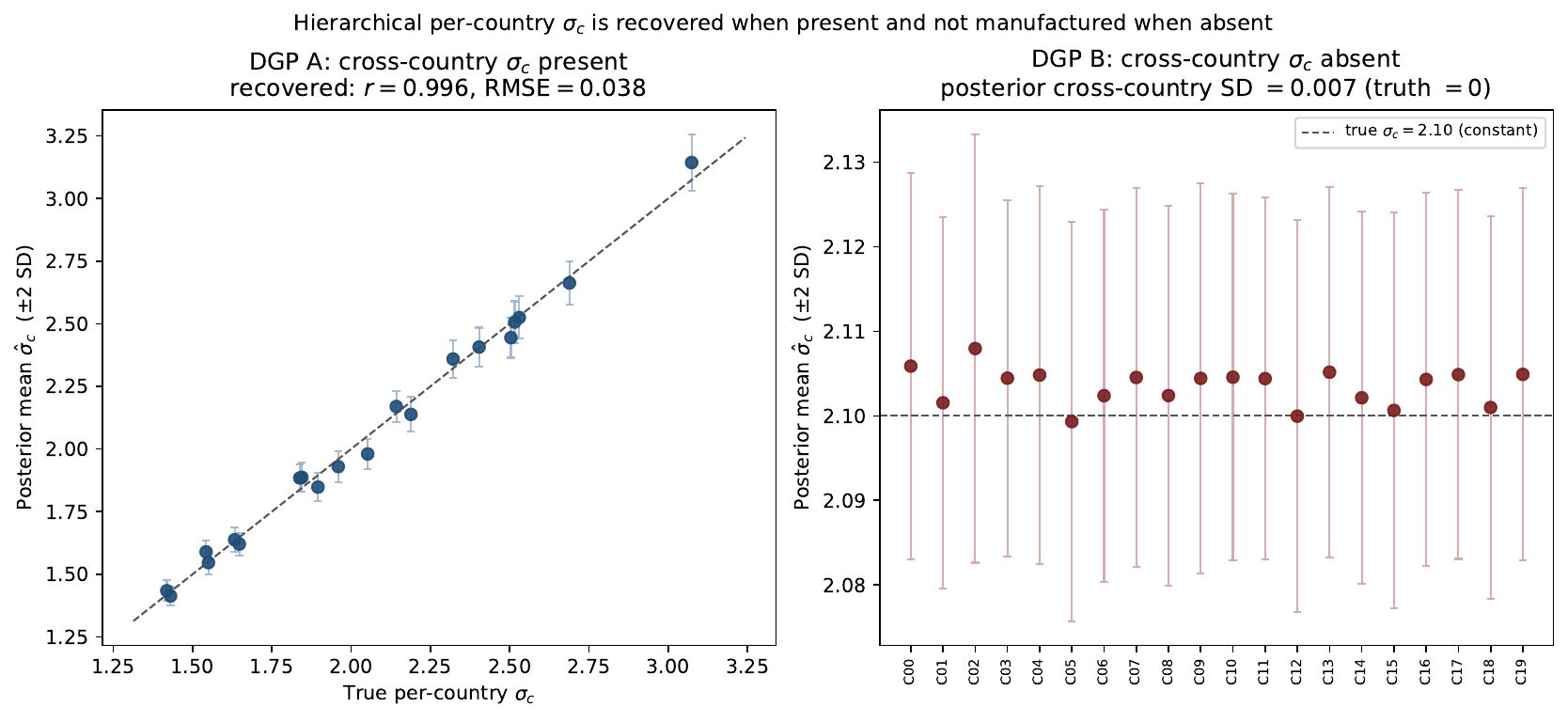}
\caption[Synthetic recovery of the per-country scale]{\textbf{Synthetic recovery of the per-country scale $\sigma_c$.} Synthetic recovery of the hierarchical per-country scale $\sigma_c$ on
a multi-group reallocation data-generating process. \textbf{(a)}~DGP~A, where the true $\sigma_c$
varies across groups: posterior means (with $\pm2$ posterior SD) against the
truth, hugging the identity line ($r=0.996$, RMSE$=0.038$). \textbf{(b)}~DGP~B,
where the true $\sigma_c$ is constant and groups differ only in \FVR{} rate: the
posterior places every group near the common truth, with cross-country
posterior SD $0.007$ (truth $0$) --- no spurious dispersion is manufactured.}
\label{fig:sigma-recovery}
\end{figure}

\paragraph{Synthetic recovery of the hierarchical scale.}
The model was validated
on
a synthetic multi-group hierarchical per-country $\log\sigma_c$ reallocation data-generating process
with a known truth. \Figref{sigma-recovery} shows  two cases. When
the true $\sigma_c$ varies across groups (DGP~A), the posterior recovers it
closely (Pearson $r = 0.996$, RMSE $= 0.038$). When the true $\sigma_c$ is
constant (DGP~B) --- with the groups differing only in their \FVR{} rates --- the
posterior does \emph{not} manufacture spurious cross-country scale variation to
absorb that heterogeneity: the posterior cross-country standard deviation of
$\sigma_c$ is $0.007$ against a truth of zero. The model thus expresses genuine
dispersion differences when they exist and reports their absence when they do
not, which is the property the cross-country inequality claims rely on.

\end{document}

%% file: ineqav-regression-table.tex
    \renewcommand{\ctNtabCols}{3}
    \renewcommand{\ctFirstHeader}{\hline
  & (1) & (2)\\ 
\hline\hline
}
    \renewcommand{\ctSubsequentHeaders}{\ctFirstHeader}
    \renewcommand{\ctBody}{Reported SD $\sigma_c^{\rm obs}$&\wrapSigOneThousandth{4.3}&\wrapSigOneThousandth{4.4}\\  
&\coefse{.026}&\coefse{.026}\\  
Inequality aversion $a$&\wrapSigOneThousandth{$-$.084}&\wrapSigOneThousandth{.33}\\  
&\coefse{.002}&\coefse{.029}\\  
$a \times \sigma_c^{\rm obs}$ ($\beta_{e\sigma}$)&&\wrapSigOneThousandth{$-$.38}\\  
&&\coefse{.026}\\  
$\log$ income&$-$.002&$-$.002\\  
&\coefse{.004}&\coefse{.004}\\  
Constant&\wrapSigOneThousandth{1.22}&\wrapSigOneThousandth{1.14}\\  
&\coefse{.029}&\coefse{.029}\\ 
\cline{1-\ctNtabCols}
 }
    
    \renewcommand{\ctCaption}{ {\color{blue}  } }
    \ifx\@ctUsingWrapper\@empty
    \begin{table}
    \begin{tabular}{ccc}
    \ctFirstHeader
    \ctBody
    \end{tabular}
    \end{table}
    \else
    \fi

    \renewcommand{\ctStartTabular}{\begin{tabular}{ccc}}
    \renewcommand{\ctStartLongtable}{\begin{longtable}[c]{ccc}}
    